%% file: main.tex
\documentclass[journal]{vgtc}                     %

\onlineid{1554}

\vgtccategory{Research}

\vgtcpapertype{Theoretical \& Empirical}

\title{A Comparative Study of the Perceptual Sensitivity of Topological Visualizations to Feature Variations}

\author{%
  Tushar~M.~Athawale, Bryan~Triana, Tanmay~Kotha, Dave~Pugmire, and~Paul~Rosen%
}

\authorfooter{
  \item
  	T. Athawale and D. Pugmire were with Oak Ridge National Laboratory.
  	E-mail: \{ athawaletm | pugmire \}@ornl.gov
  \item
  	B. Triana and T. Kotha were with the University of South Florida.

  \item P. Rosen was with the University of Utah.
  	E-mail: paul.rosen@utah.edu
      
}

\abstract{%
\input{sec-abstract}
}

\keywords{Perception \& cognition, computational topology-based techniques, comparison and similarity.}

\teaser{
    \centering
    \includegraphics[width=0.82\linewidth]{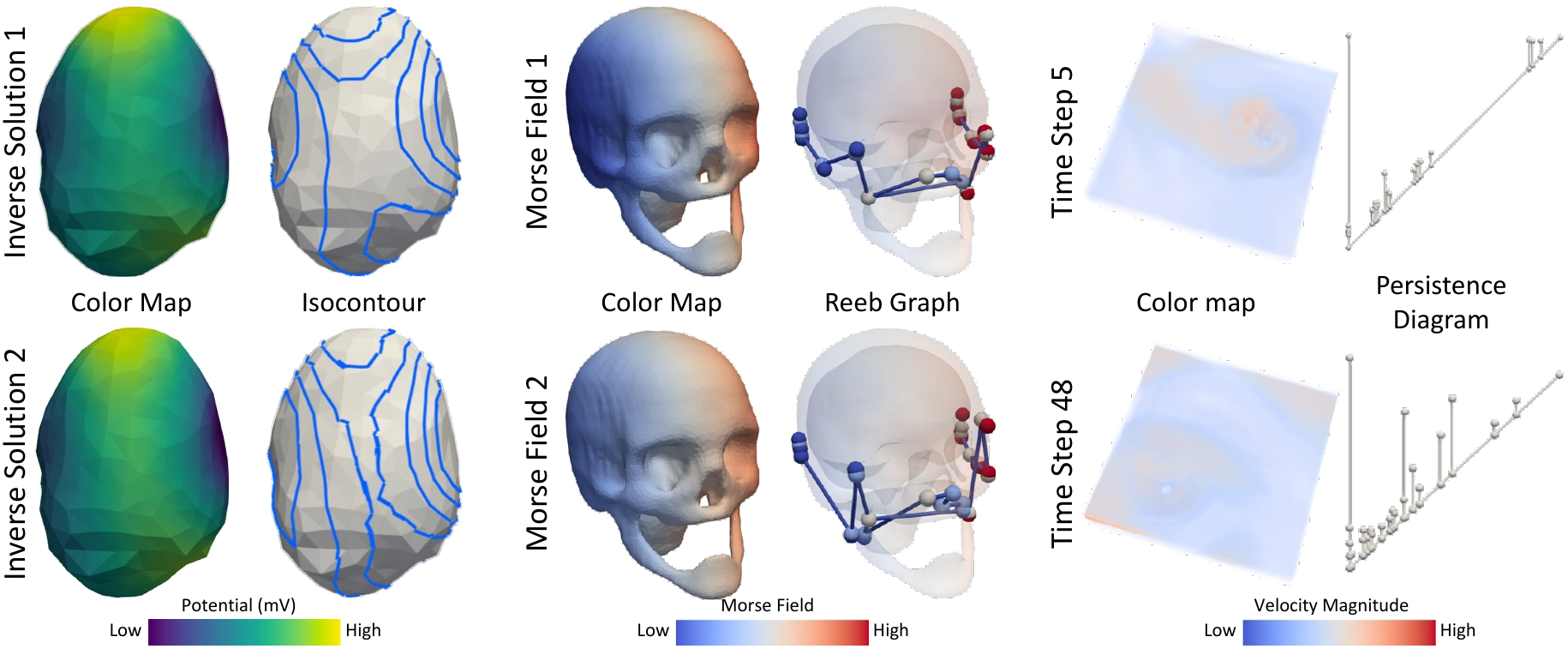}

    \vspace{-11pt}
    \subfloat[Potentials on heart surface~\cite{TA:2022:Njeru:ECGIuncertainty}\label{fig:topoVisComparisonRealData:njeru}]{\hspace{0.25\linewidth}}
    \hspace{0.0\linewidth}
    \subfloat[Patient-specific cephalometric landmarks~\cite{makram2014reeb}\label{fig:topoVisComparisonRealData:makram}]{\hspace{0.325\linewidth}}
    \subfloat[Time-varying simulations of a hurricane~\cite{TA:2020:Vidal:persistenceDiagramBarycenter}\label{fig:topoVisComparisonRealData:vidal}]{\hspace{0.28\linewidth}}

    \vspace{-9pt}
    \caption{Recreated examples of visual comparisons of scalar field data from prior work. Comparing states requires identifying changes in the position and amplitude of the various data features. In these examples, the left images are the color mapped versions of the data, while the right are various forms of topology-based visualizations that explicitly communicate the underlying topological structures.} %
    \label{fig:topoVisComparisonRealData}
}

\graphicspath{{figs/}{figures/}{pictures/}{images/}{./}} %

\usepackage{tabu}                      %
\usepackage{booktabs}                  %
\usepackage{lipsum}                    %
\usepackage{mwe}                       %

\usepackage{mathptmx}                  %

\usepackage{amsfonts}
\usepackage{comment}
\usepackage{multicol}
\usepackage{multirow}
\usepackage{amsmath}
\usepackage{bm}
\usepackage{setspace}

\newcommand{\Rspace}        {{\mathbb R}}

\newcommand {\mm}[1] {\ifmmode{#1}\else{\mbox{\(#1\)}}\fi}

\newcommand{\Mgroup}        {{\mathcal M}}

\newcommand{\para}[1]{\textbf{#1:}}

\renewcommand{\paragraph}[1]{\textbf{#1:}}

\newcommand\blfootnote[1]{%
  \begingroup
  \renewcommand\thefootnote{}\footnote{#1}%
  \addtocounter{footnote}{-1}%
  \endgroup
}

\begin{document}

\firstsection{Introduction}

\maketitle

\setstretch{0.95}

\input{sec-intro}

\input{sec-prior}

\input{sec-background}

\input{sec-visualizations}

\input{sec-method}

\input{sec-exp_env}

\setstretch{0.947}

\input{sec-experiment}

\input{sec-results}

\input{sec-discussion}

\setstretch{0.98}
\acknowledgments{%
This work was supported in part by a grant from the National Science Foundation (III-2316496) and by the U.S. Department of Energy (DOE) RAPIDS-2
SciDAC project under contract number DE-AC05-00OR22725.
}

\bibliographystyle{abbrv-doi-hyperref}

\bibliography{main}

\end{document}

%% file: sec-abstract.tex
Color maps are a commonly used visualization technique in which data are mapped to optical properties, e.g., color or opacity. Color maps, however, do not explicitly convey structures (e.g., positions and scale of features) within data. Topology-based visualizations reveal and explicitly communicate structures underlying data. Although we have a good understanding of what types of features are captured by topological visualizations, our understanding of people's perception of those features is not. This paper evaluates the sensitivity of topology-based isocontour, Reeb graph, and persistence diagram visualizations compared to a reference color map visualization for synthetically generated scalar fields on 2-manifold triangular meshes embedded in 3D. In particular, we built and ran a human-subject study that evaluated the perception of data features characterized by Gaussian signals and measured how effectively each visualization technique portrays variations of data features arising from the position and amplitude variation of a mixture of Gaussians. For positional feature variations, the results showed that only the Reeb graph visualization had high sensitivity. For amplitude feature variations, persistence diagrams and color maps demonstrated the highest sensitivity, whereas isocontours showed only weak sensitivity. These results take an important step toward understanding which topology-based tools are best for various data and task scenarios and their effectiveness in conveying topological variations as compared to conventional color mapping.

%% file: sec-intro.tex
The scale and complexity of scientific datasets have reached a level that makes directly communicating the details of the data through visualization exceedingly difficult~\cite{TA:Hansen:2014:sciVis}. Even if such data can be rendered, the complexity of the output far exceeds what humans can directly interpret~\cite{Tory:2004,varakin2004unseen}. One approach to resolving this issue is to use topological data analysis~(TDA) to summarize the data~\cite{TA:Tierny:2017:TDA, TA:Yan:2021:ScalarFieldComparisonTopoDescriptors}. TDA shows promise in this area because it provides a suite of tools that can summarize $n$-dimensional scalar field data as features or hierarchies in ways that are intuitive to humans~\cite{wong_gestalt_2010,quadri2020modeling}. TDA-based visualizations have been fundamental to understanding feature variations in multiple real-world applications, including fluid dynamics~\cite{TA:Favelier:2019:criticalPointVariabilityEnsembles, TA:2022:Nauleau:turbulentFlowTopoAnalysis} and combustion~\cite{TA:2010:Bremer:combustionTopo}. Although what data features TDA can summarize is quite well understood (e.g., the relationship of critical points of a scalar function), \textit{what TDA-based visualizations communicate to a user is not}. 
In this paper, we describe our study of the sensitivity of TDA-based visualization methods, namely, isocontours, Reeb graphs, and persistence diagrams, to a reference color mapping visualization in communicating feature variations within a scalar field.
\blfootnote{This manuscript has been authored by UT-Battelle, LLC under Contract No. DE-AC05-00OR22725 with the U.S. Department of Energy. The publisher, by accepting the article for publication, acknowledges that the U.S. Government retains a non-exclusive, paid up, irrevocable, world-wide license to publish or reproduce the published form of the manuscript, or allow others to do so, for U.S. Government purposes. The DOE will provide public access to these results in accordance with the DOE Public Access Plan (\url{http://energy.gov/downloads/doe-public-access-plan}).}

\begin{figure}[!ht]
    \centering

    \hspace{40pt}
    \subfloat[Color Map\label{fig:teaser:a}]{\hspace{0.13\linewidth}}
    \hspace{15pt}
    \subfloat[Isocontour\label{fig:teaser:b}]{\hspace{0.13\linewidth}}
    \hspace{15pt}
    \subfloat[Reeb Graph\label{fig:teaser:c}]{\hspace{0.14\linewidth}}
    \hspace{5pt}
    \subfloat[Persistence Diagram\label{fig:teaser:d}]{\hspace{0.23\linewidth}}
    
    \vspace{-5pt}
    \includegraphics[trim=0 0 0 16px, clip, width=0.875\linewidth]{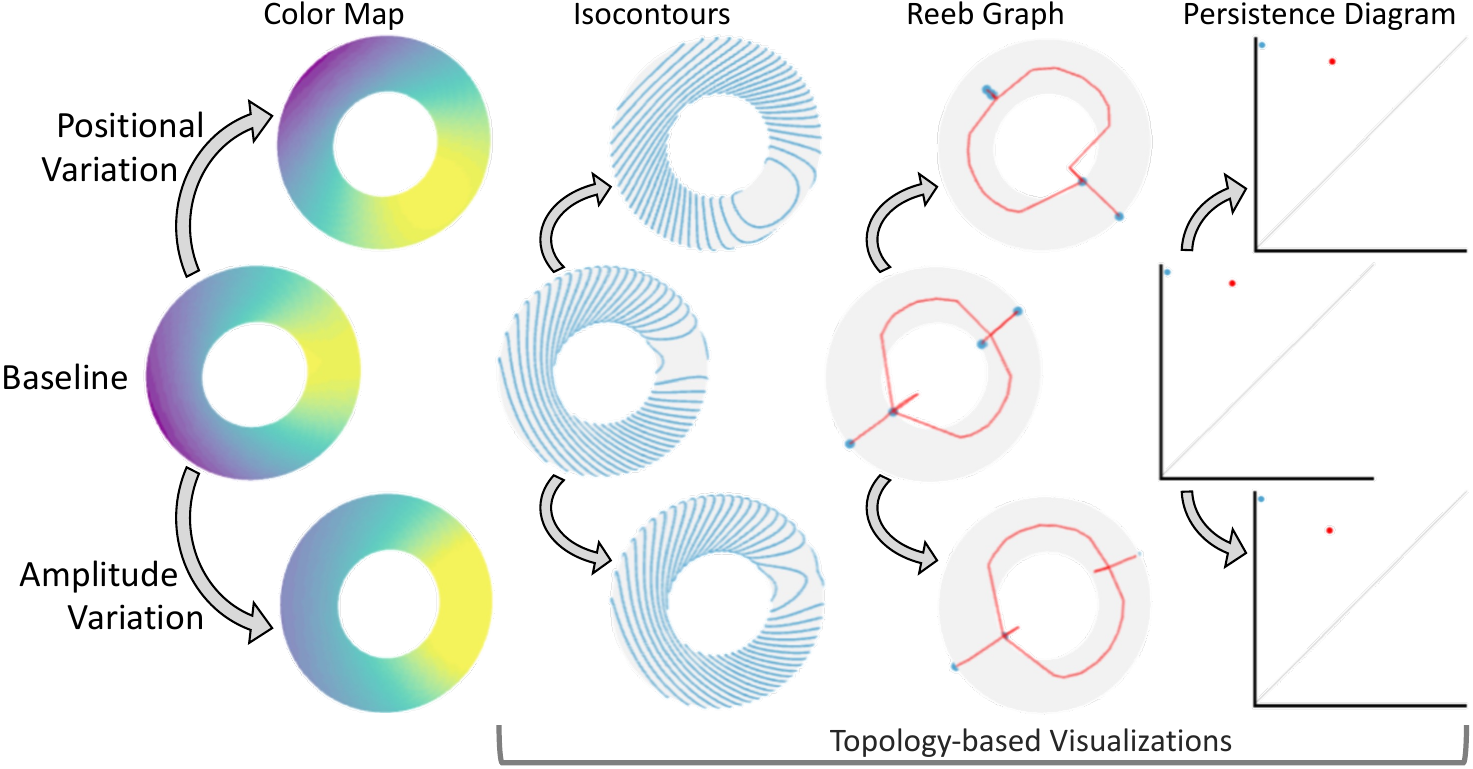}

    \vspace{-5pt}
    \caption{Demonstration of the sensitivity of the visualizations to variations of features compared with a baseline (middle). The positional variations (top) are reflected via the change in the position of hot spots in the color map and the orientation of features in isocontours and Reeb graphs, but they do not affect the persistence diagram. Amplitude variations (bottom) are reflected in the brightness in the color map, the distribution of lines in isocontours, and the movement of dots in the persistence diagram, but they do not prominently impact Reeb graphs. Our experimental stimuli (see supplement) contain multiple features and noise.}
    \vspace{-8pt}
   \label{fig:teaser}    
   
\end{figure}

In practice, color mapping (see \autoref{fig:teaser:a}) is often used to directly render the scalar field, but such an approach leaves the function topology to be inferred by the viewer. 
On the other hand, TDA-based visualizations show abstractions of the scalar field topology.
The isocontour visualization (see \autoref{fig:teaser:b}) acts as a topographic map of the function by using a series of level sets that directly show changes in the topology of the function. Reeb graphs (see \autoref{fig:teaser:c}) act as a skeletal summary of the isocontour visualization by tracking the evolution, i.e., creation, destruction, merging, and splitting, of the contour level sets and specifically highlight critical points in the function. Finally, persistence diagrams  (see \autoref{fig:teaser:d}) abstract the topology of a Reeb graph by pairing critical points and presenting them in a scatterplot-like display. Each visualization has advantages and disadvantages in terms of intuitiveness, visual clutter, and the types of insights they purport to show.

These visualizations are frequently employed for comparing datasets in scientific applications. For example, Njeru et al.~\cite{TA:2022:Njeru:ECGIuncertainty} compared inverse solutions for potentials on the heart surface through isocontours to gain insight into positional uncertainty of the source of arrhythmia (see \autoref{fig:topoVisComparisonRealData:njeru}). Makram and Kamel~\cite{makram2014reeb} analyzed Reeb graphs of Morse functions mapped to the human skull to extract and compare patient-specific cephalometric landmarks (see \autoref{fig:topoVisComparisonRealData:makram}). Finally, Vidal et al.~\cite{TA:2020:Vidal:persistenceDiagramBarycenter} compared persistence diagrams of time-varying datasets (see \autoref{fig:topoVisComparisonRealData:vidal}) and developed a novel technique to compute the barycenter of persistence diagrams that is visually closer to individual time steps.

This paper provides a first-of-its-kind empirical evaluation of the efficacy of isocontours, Reeb graphs, and persistence diagrams against a reference of color mapping in performing a visual comparison task for scalar fields. 
Specifically, our evaluation looks at scalar fields defined on 2-manifold triangular mesh with no boundary, embedded in 3D that can be rotated and zoomed (similar to \autoref{fig:topoVisComparisonRealData:njeru} and \autoref{fig:topoVisComparisonRealData:makram}). Further, the scalar fields are modeled as a mixture of a  small number of Gaussian signals (a commonly used data model in scientific analysis, e.g., see~\cite{TA:2020:Vidal:persistenceDiagramBarycenter, TA:2020:Yan:mergeTreeAverage}), where {\em amplitude} and {\em position} refer to the peak and mean of a Gaussian, respectively (see \autoref{sec:mathematicalModel} for more details). 
    Since the amplitude and position of Gaussian signals are building blocks of our data model, by considering those two types of variations, we can gain a better sense of the saliency of information within these visualizations. 
    To do this, we created and conducted a human-subject study involving 102 non-expert participants who performed a comparison task on these visualizations. Our high-level findings are:
\begin{itemize}[noitemsep]
\item Our study confirmed some of our expectations for these visualizations. For instance, Reeb graphs were reasonably sensitive to positional variations, whereas persistence diagrams and color maps showed high sensitivity to amplitude variations. 
\item Some counterintuitive results also surprised us. For instance, color maps showed sensitivity to amplitude variations but not positional variations, and isocontours showed no sensitivity to positional variations and only weak sensitivity to amplitude variations. 
\item The results of this study provide an important step toward understanding the communication of topological features in these commonly utilized visualization techniques. Furthermore, our results provide important insights into the limitations of these visualizations and guidance on how techniques and systems can be improved to provide more reliable insights into data.
\end{itemize}

%% file: sec-prior.tex
\section{Prior Work}
We briefly discuss related research, including the evaluation of visualizations, topology-based data visualization, and sensitivity analysis.

\subsection{Evaluation of Visualization}

Although there are many approaches to visualizing complex scientific data, measuring how good a particular visualization is has been a nontrivial challenge for researchers~\cite{TA:Johnson:2004:topSciVisProblems}. Visualization quality is determined by the ability of users to effectively and efficiently extract knowledge from complex, large-scale data. Various visualization components (e.g., hardware requirements, software implementation costs, interactivity, accuracy of visualizations, and perception and cognition) have been evaluated to understand the overall impact of visualizations on decision-making~\cite{TA:Wijk:2005:valueOfVis, TA:Fekete:2008:valueOfInfovis, TA:Munzner:2009:nestedModelVisDesignAndValidation, quadri2021survey}. One promising approach to evaluating the effectiveness of visualizations in decision-making is conducting human-subject studies. Human-subject studies enable researchers to perform quantitative assessment of visualization parameters (e.g., errors and time associated with decisions, comparison with alternative visualizations, and the sensitivity of decisions to a person's domain and visualization expertise), which can potentially help reinforce the scientific foundation of visualization~\cite{TA:Kosara:2003:userStudiesVisViewPoints, TA:North:2006:visInsightAnalysis}. 

Multiple studies have assessed the effectiveness of various scientific data visualizations. Color mapping is one of the fundamental methods used in data visualization because scalar data can be encoded into intuitive visual attributes, such as hue and opacity, to convey data features. The choice of a color map, however, strongly influences human perception~\cite{TA:Moreland:2009:divergingColormaps, TA:Zhou:2019:colormapVis}. Liu and Heer~\cite{viridis_colormap} quantified the perceptual performance of users for different color maps by presenting tasks that required color comparisons. A similar study was performed to understand the effectiveness of 2D~\cite{TA:Laidlaw:2005:2dvectorFieldVisUserStudy} and 3D~\cite{TA:Forsberg:2009:3dvectorFieldVisUserStudy} vector field visualizations by evaluating the quality of decisions in identifying flow features (e.g., critical points and direction of particle advection). 
An interesting approach to measuring cognitive load experienced when perceiving information from visualizations is through the analysis of EEG patterns of subjects~\cite{TA:Anderson:2011:visEffectivenessEEG}. %

\subsection{Topology-Based Data Visualization}

As a key tool in scientific visualization, topological data analysis (TDA)~\cite{TA:Hauser:2007:TDA, TA:Tierny:2017:TDA, TA:Yan:2021:ScalarFieldComparisonTopoDescriptors} enables the understanding of abstract structures present in data. Visualization applications in numerous scientific domains (e.g., molecular dynamics~\cite{TA:Natarajan:2008:molecularSurfaceVisTDA}, combustion science~\cite{TA:Bremer:2010:burningStructuresTDA}, fluid dynamics~\cite{TA:Laney:2006:mixingFluidsTDA}, and radio astronomy~\cite{rosen2021using}) have demonstrated the strengths of TDA in unraveling complexities of the underlying data. A few of the fundamental topological descriptors for univariate data include level sets~\cite{Lorensen:1987:MCA}, critical points~\cite{TA:Morse:1930:criticalPoints}, Reeb graphs~\cite{computational_topology_intro}, contour trees~\cite{compute_contour_trees}, and persistence diagrams~\cite{persplot_stability} (see \autoref{cp:bg} for technical details on these descriptors). Other topological descriptors include Morse-Smale complexes~\cite{TA:Edelsbrunner:2003:MorseSmaleComplexes}, which are widely used topological abstractions that segment the domain into cells with uniform gradient behavior. Reeb spaces~\cite{TA:Edelsbrunner:2008:reebSpaces}, Jacobi sets~\cite{TA:Tierny:2017:jacobiSetsReebSpaces}, and fiber surfaces/feature level sets~\cite{TA:Carr:2015:fiberSurfaces, TA:Jankowai:2020:featureLevelSets} extend the concepts of Reeb graphs, critical points, and level sets, respectively, to multivariate data. Furthermore, there have been significant research developments in leveraging topological features for effective feature tracking of scientific data~\cite{TA:Tricoche:2002:vectorTensorFeatureTrackingTDA, TA:2003:Post:flowFeatureTrackingSurvey, TA:Sohn:2006:contourTopologyTracking, TA:Bremer:2007:TDAfeatureTracking, TA:Chen:2013:fluidParticleTracking, TA:Bujack:2020:timeDependentFlowTopo, TA:Wito:2021:featureTrackingTDACyclones}.

\begin{figure*}[!t]
    \centering

    \begin{minipage}[b]{0.535\linewidth}
        \fbox{\includegraphics[height=3.41cm]{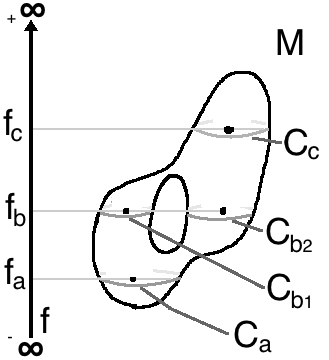}}
        \hspace{1pt}
        \fbox{\includegraphics[height=3.41cm]{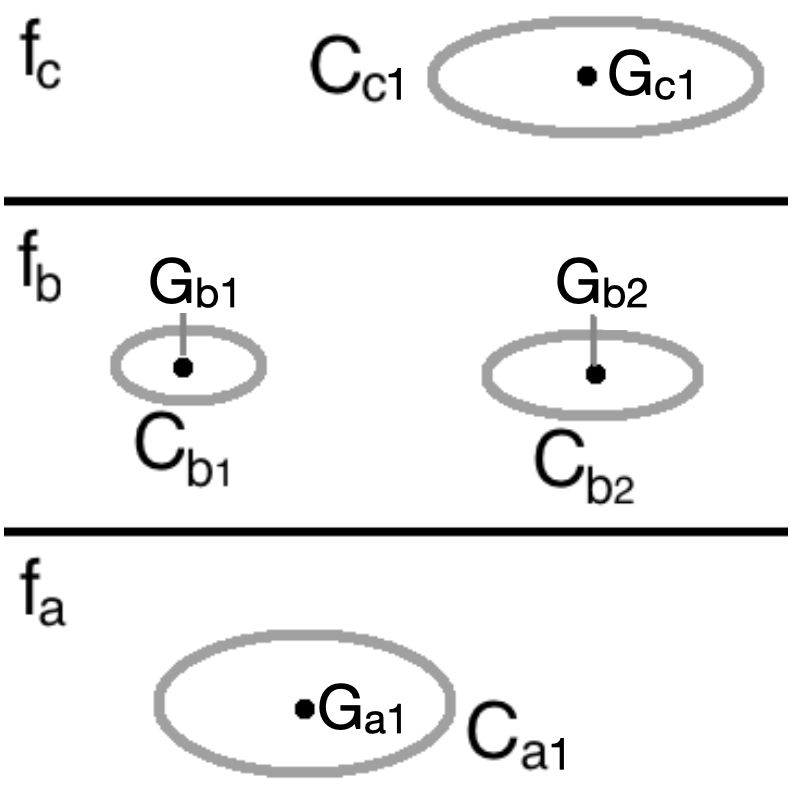}}
        \hspace{1pt}
        \fbox{\includegraphics[height=3.41cm]{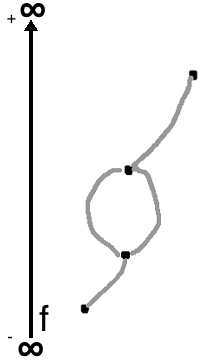}}
    \end{minipage}
    \begin{minipage}[b]{1pt}
        \hspace{-193pt}
        \subfloat[\label{fig:reeb_graph1:sub1}]{}
        \hspace{106pt}
        \subfloat[\label{fig:reeb_graph1:sub2}]{}
        \hspace{67pt}
        \subfloat[\label{fig:reeb_graph1:sub3}]{}
        \vspace{83pt}
    \end{minipage}
    \hspace{-2pt}
    \rule[10pt]{1pt}{2.9cm}
    \hspace{5pt}%
    \begin{minipage}[b]{0.225\linewidth}
        \fbox{\includegraphics[width=0.42\linewidth]{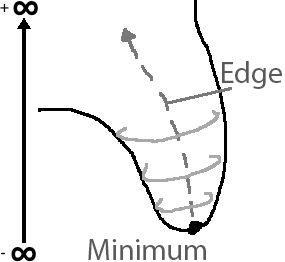}}
        \hspace{1pt}
        \fbox{\includegraphics[width=0.42\linewidth]{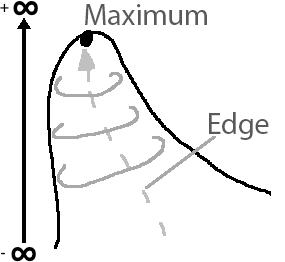}}

        \fbox{\includegraphics[width=0.42\linewidth]{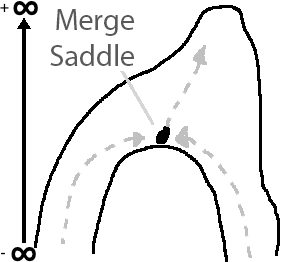}}
        \hspace{1pt}
        \fbox{\includegraphics[width=0.42\linewidth]{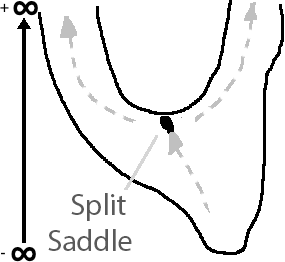}}
    \end{minipage}
    \begin{minipage}[b]{1pt}
        \hspace{-15pt}
        \colorbox{White}{\hspace{4pt}}
        \vspace{39pt}
    \end{minipage}
    \begin{minipage}[b]{1pt}
        \hspace{-77pt}
        \subfloat[\label{fig:crit_pts_types:a}]{}
        \hspace{58pt}
        \subfloat[\label{fig:crit_pts_types:b}]{}
        
        \vspace{32pt}
        \hspace{-76pt}
        \subfloat[\label{fig:crit_pts_types:c}]{}
        \hspace{57pt}
        \subfloat[\label{fig:crit_pts_types:d}]{}
        \vspace{31pt}
    \end{minipage}
    \hspace{-2pt}
    \rule[10pt]{1pt}{2.9cm}
    \hspace{5pt}%
    \begin{minipage}[b]{0.106\linewidth}
        \fbox{\includegraphics[width=\linewidth]{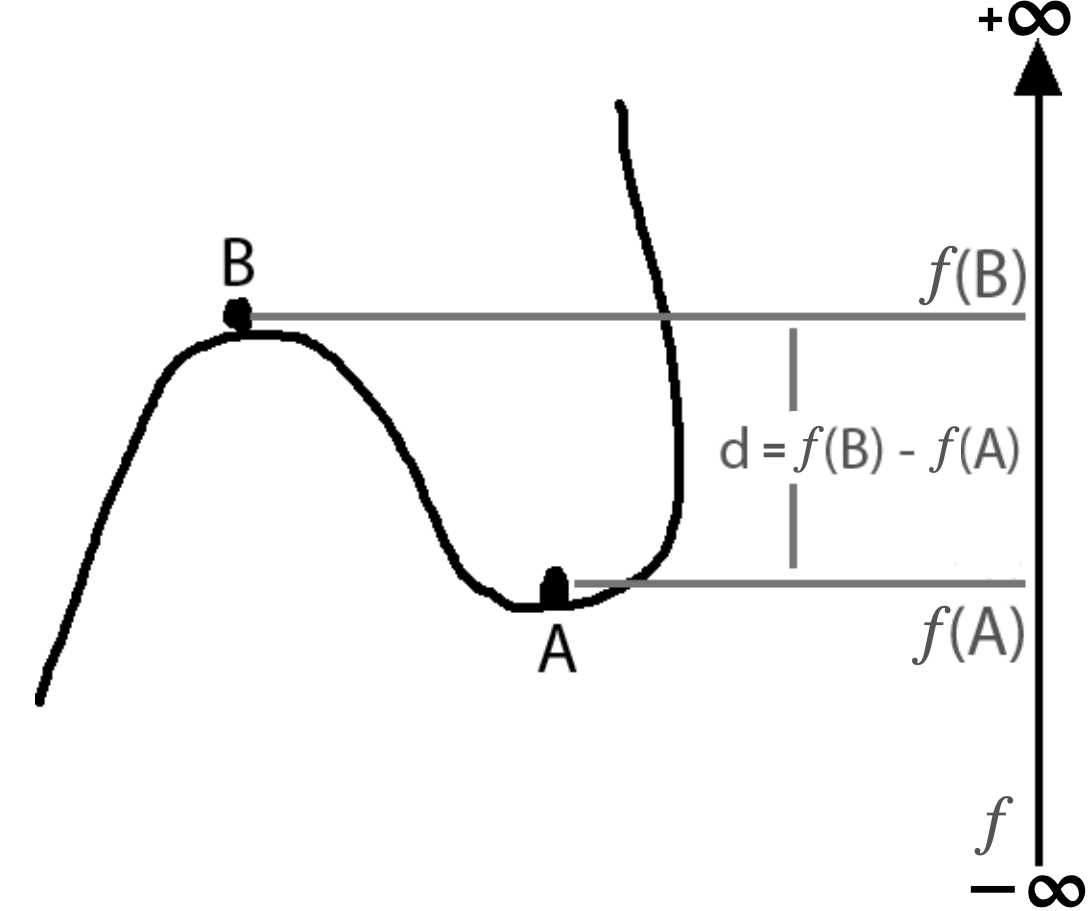}}

        \fbox{\includegraphics[width=\linewidth]{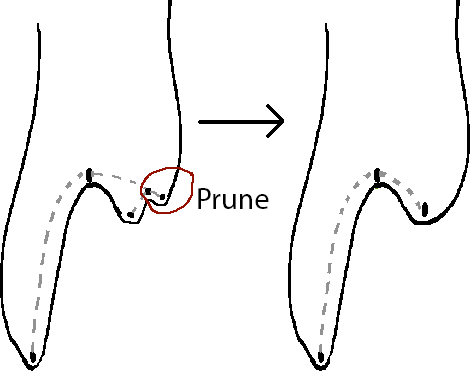}}
    \end{minipage}
    \begin{minipage}[b]{1pt}
        \hspace{-58pt}
        \colorbox{White}{\hspace{4pt}}
        \vspace{37pt}
    \end{minipage}
    \begin{minipage}[b]{1pt}
        \hspace{-60pt}
        \subfloat[\label{fig:distance_d_definition}]{}

        \vspace{33pt}
        \hspace{-60pt}
        \subfloat[\label{fig:prune_example}]{}
        \vspace{30pt}        
    \end{minipage}
    
    \vspace{-3pt}
    \caption{Reeb graph construction and critical point types are described. Left Panel: 
    (a)~The isolevels, $f_z = (f_a, f_b, f_c)$, of function, $f$, derive level sets, $C_z = (C_a, C_b, C_c)$.
    (b)~The number of connected components of a level set for each function value depends on the critical points of the function. Both $f_a$ and $f_c$ have a level set with one connected component, whereas $f_b$ has two. The center point for each connected component, $C_{zi}$, is labeled as $G_{zi}$. (c)~By tracking the points, $G_{zi}$, from $-\infty\rightarrow+\infty$, the Reeb graph structure emerges, which summarizes the topology succinctly.
    Center Panel: The four types of critical points, including 
    (d)~a local minimum in the function, which has one outgoing edge; 
    (e)~a local maximum, which has one incoming edge; 
    (f)~the merge saddle, which has two incoming edges that merge into a single outgoing edge; and (g)~the split saddle, which has one edge split into two.
    Right Panel: 
    (h)~For two paired critical points, $A$ and $B$, the persistence between them is $d=|f(B)-f(A)|$, where $f$ is the function applied to the manifold.
    (i)~To reduce visual clutter, the Reeb graph can be pruned by removing any birth-death pair with $d<\epsilon$.}
    \label{fig:reebgraph_description}
    \vspace{-5pt}
    
\end{figure*}

Previous works have also investigated novel metrics to quantitatively compare scalar fields via topological descriptors. For example, the distance between persistence diagrams can be measured as the cost of matching features between diagrams, e.g., using bottleneck~\cite{TA:2007:Cohen-Steiner:bottleneckDistance} or Wasserstein~\cite{computational_topology_intro} distance. Further, kernel-based methods~\cite{TA:2015:Reininghaus:kernelDistance, TA:2017:Carriere:kernelDistance} have been developed to quantify the distance between persistence diagrams that are more suitable for machine learning tasks. An array of metrics have been devised, including interleaving distance~\cite{TA:2013:Morozov:interleavingDistance}, edit distance~\cite{TA:2020:Sridharamurthy:editDistance}, stable distance~\cite{TA:2023:Bollen:stableDistance}, and Wasserstein distance~\cite{TA:2022:Pont:wassersteinDistance}, that quantitatively measure the distance between merge trees, the building blocks of contour trees and Reeb graphs. Bauer et al.~\cite{TA:2014:Bauer:reebGraphDistance} proposed functional distortion metric, and Lan et al.~\cite{TA:2023:Lan:labledInterleavingDistanceReebGraph} levereged the idea of intrinsic interleaving distances between merge trees~\cite{TA:2022:Gasparovic:intrinsicInterleavingDistance} to compare Reeb graphs. The recent survey paper by Yan et al.~\cite{TA:Yan:2021:ScalarFieldComparisonTopoDescriptors} provides an overview of the state of the art in quantitative measures for comparing topological descriptors of scalar fields.

\subsection{Sensitivity Analysis} 

Since we evaluate the sensitivity of visualizations to amplitude and positional variations of scalar fields, we discuss how sensitivity analysis has been shown to help improve data representations and scientific analyses. Saltelli et al.~\cite{saltelli2008global} discussed in detail a broad spectrum of statistical techniques for understanding the sensitivity of functions to their input parameters. Cacuci et al.~\cite{cacuci2003sensitivity, cacuci2005sensitivity} provided the theoretical foundation for sensitivity and analyzed sensitivity in various applications, including fluid dynamics and atmospheric science. In the context of visualization, Brecheisen et al.~\cite{Brecheisen:2009:DTIsensitivity} analyzed the sensitivity of fiber tracking of diffusion tensor imaging to input parameters. 
Chan et al.~\cite{Chan:2010:scatterPlotUncertainty, Chan:2013:scatterPlotUncertainty} proposed techniques to quantify the local sensitivity of high-dimensional data and encode such sensitivity into 2D scatterplots for improved data analyses. Finally, Liu et al.~\cite{Liu:2014:sensitivityDrivenTF} encoded the sensitivity of direct volume rendering into the transfer function space to help users efficiently and effectively design transfer functions.

%% file: sec-background.tex
\section{Background on TDA}
\label{cp:bg}
Here, we briefly present technical definitions of level sets, Reeb graphs, and persistence diagrams, which we utilized in our study.

\subsection{Contour Level Sets and Sublevel Sets} 
\label{sec:levelSets}
Level sets are a fundamental tool for analyzing scientific data. 
Let $f:\Mgroup\rightarrow \Rspace$ be a scalar function defined on an $m$-dimensional manifold, $\Mgroup$. A level set, $C$, of the function, $f$, for the isolevel, $f_z$, corresponds to a pre-image of function value, $f_z$. Mathematically, $C(f_z)\equiv\{P: P\in \Mgroup \land f(P)=f_z\}$, where $P$ denotes domain positions. In a related notion, the sublevel set, $\mathcal{L}$, of function, $f$, for isovalue $f_z \in \Rspace$  is $\mathcal{L}(f_z)\equiv\{ P \in \Mgroup| f(P) \le f_z\}$. The visualization of level sets, called \textit{isocontouring}, is discussed in \autoref{sec:visualization:iso}.

\subsection{The Reeb Graph} 
\label{sec:reebGraphs}
The Reeb graph is a structural abstraction that provides insight into the topological skeleton of scalar field data. Formally, for an $m$-dimensional manifold, $\Mgroup$ (which has a Morse function, $f$, mapped over the surface, $f:\Mgroup\rightarrow \Rspace$), the Reeb graph tracks the connected components of the level sets as $f$ is swept from $-\infty \rightarrow +\infty$~\cite{computational_topology_intro}. The Reeb graph of a Morse function, $f$, defined on a simply-connected manifold, $\Mgroup$, is loop-free and is referred to as a contour tree~\cite{carr2003computing}. \autoref{fig:reebgraph_description}(left panel) summarizes how connected components of level sets evolve as the iso-level, $f_z$, is swept. In \autoref{fig:reeb_graph1:sub1}, each connected component of the level set for iso-level, $f_z$, is denoted by $C_{zi}$ with $i \in N$, where $N$ denotes the number of connected components. Each connected component, $C_{zi}$, can be collapsed to a single point, $G_{zi}$, as illustrated in \autoref{fig:reeb_graph1:sub2}. The collapsed points, $G_{zi}$, are connected to derive the Reeb graph in \autoref{fig:reeb_graph1:sub3}.

Tracking the evolution of connected components, $C_{zi}$, special events change the number, $N$, of connected components. These events are associated with critical points in the topology of $f$ and are positions on the manifold, where $\nabla f = 0$ (see \autoref{fig:reebgraph_description}(center panel)). In the case of a local minimum, a new edge begins (see \autoref{fig:crit_pts_types:a}), while an edge ends for a local maximum (see \autoref{fig:crit_pts_types:b}). In the case of saddle points, two or more edges can merge into one (see \autoref{fig:crit_pts_types:c}), or one edge can split into multiple branches (see \autoref{fig:crit_pts_types:d})~\cite{parallel_reeb}. Therefore, the structure of the Reeb graphs is determined by the critical points of the scalar field. We calculate the Reeb graph using Recon~\cite{doraiswamy2012computing}.

\subsection{Persistence Diagrams}
\label{sec:pd}
The structure of the Reeb graph can be summarized with a multiset of points, known as a \textit{persistence diagram}. Persistence diagrams highlight the more prominent features in a Reeb graph and are a stable representation of the function~\cite{computational_topology_intro,persplot_stability}. 

Points of the persistence diagram are formed by pairing critical points in one of three configurations, saddle-minimum, saddle-maximum, and saddle-saddle pairs. Without a loss of generality, we describe saddle-minimum pairing, which depends on the sublevel sets of the function and corresponds to the Elder's rule~\cite{computational_topology_intro}.  As $f_z$ increases continuously from $-\infty \rightarrow +\infty$, let $B$ be a saddle point of the function with function value $f_z=f(B)$. Let $A$ be the last unpaired minimum added to the sublevel set component of B, $\mathcal{L}(f_z=f(B))$. $A$ and $B$ are grouped together as a \textit{birth-death pair} feature, where $A$ marks the birth point of the feature, and $B$ marks the death point of the feature. 

Of the three types of birth-death pairs that exist in Reeb graphs, saddle-minimum and saddle-maximum pairs are identified using an approach called \textit{branch decomposition}~\cite{pascucci2004multi}, whereas saddle-saddle pairs, formed by tunnels in the manifold, are identified using an extension of branch decomposition~\cite{tu2019propagate}.
A measure, known as \textit{persistence}, $d=|f(B) - f(A)|$, is applied to the pair, as shown in \autoref{fig:distance_d_definition}.

Persistence is used to classify the importance of a feature and can be used to differentiate topological signal from noise. One method of denoising a Reeb graph is to prune noise via \textit{persistence simplification}~\cite{TA:Edelsbrunner:2002:persistenceSimplification}, as illustrated in \autoref{fig:prune_example}. For a given feature to be considered noise, its persistence, $d$, must be less than a certain $\epsilon$ threshold. The features are removed from the output graph by deleting associated nodes and reconnecting the graph~\cite{ReebGraphSimplification}. Selecting an optimal threshold is difficult to automate because being too aggressive can lead to the removal of important features, and being too relaxed may result in a noisy dataset. Therefore, users of persistence often rely on manual selection and tuning of $\epsilon$ to gain insights from Reeb graphs.

%% file: sec-visualizations.tex
\input{fig.vis.evaluated}

\section{Visualizations Evaluated}
\label{sec:visualization}
We compare a reference visualization type---color maps---to three TDA-based visualizations---isocontours, Reeb graphs, and persistence diagrams---that reveal the topology of a function applied to a manifold. %

\subsection{Reference Visualization: Color Maps}
\label{sec:visualization:colormap}

Color maps
are a commonly used visualization and are seen as an intuitive way to interpret data (see \autoref{fig:vis_examples:color})~\cite{TA:Moreland:2009:divergingColormaps, TA:Zhou:2019:colormapVis}.
Given a function, $f$, applied to the surface of a manifold, $\Mgroup$, with a global minimum of $f_{min}$ and a global maximum of $f_{max}$\footnote[1]{$f_{min}$/$f_{max}$ are the global min/max of all three datasets in our experiments.}, a color scale, $K$, is a set of colors mapped between $K_{min}$ and $K_{max}$, respectively. 
Practically, color maps are implemented with a discrete set of colors, and in general, a given input, $f_i$, is linearly interpolated to convey the continuity of the function. However, multi-hue color maps, e.g., viridis, are formed by tracing curves through perceptually-uniform color models, e.g., CIELAB~\cite{TA:CID:1977:CIELAB} and CAM02-UCS~\cite{TA:LUO:2006:CIECAM}.
\autoref{fig:vis_examples:color} shows example visualizations
with $f_{min}$ mapped to a purple hue and $f_{max}$ mapped to a yellow hue.

\paragraph{Design} One challenge for color map visualizations is selecting the best color map to use.
For example, the size of physical marks used varies what the user perceives~\cite{modeling_color_difference}. 
Further, perceived color differences are not necessarily uniform---colors may be mathematically equidistant, while perception is biased (e.g., pure green may appear brighter than pure blue)~\cite{viridis_colormap,modeling_color_difference}. In a recent study, Liu et al.~\cite{viridis_colormap} compared various color maps and concluded that the \textit{viridis map} was the most effective at presenting data in a way that enabled users to ascertain features correctly. Cooper et al.~\cite{TA:2021:Cooper:overTheRainbox} presented a list of color maps that maintain better perceptual ordering than the rainbow color map, among which the viridis was a candidate. The luminance component, which carries magnitude information in human vision, of viridis monotonically increases (not strictly proportional) with increased data value. Moreover, for the data with low spatial frequency, the changes in saturation and hue of viridis are more effective than a grayscale color map~\cite{TA:1996:Rogowitz:colormapVis}. Since we model prominent data features as a mixture of Gaussian distributions with low spatial frequency (see~\autoref{sec:exp:data}), the viridis color map is a reasonable choice for a reference implementation. 

\paragraph{Interpretation} When observing a color map visualization, the user should look for cold spots (purple hue) and hot spots (yellow hue) to understand the positions of local minima and maxima (critical points) of the scalar field. \autoref{fig:vis_examples:color} shows cold and hot spots on the rabbit model. 

\paragraph{Interaction} Interactive rotation and zooming help overcome occlusion and enable investigating the data in finer detail.

\subsection{Topology-Based Visualization: Isocontours}
\label{sec:visualization:iso}
The visualization of level sets for different isovalues (see \autoref{sec:levelSets}), called isocontouring, provides insight into the structural evolution of a function as well as the distribution of isolevels across the manifold on which the data are sampled (see \autoref{fig:vis_examples:iso}).

\paragraph{Design} Let $f_{min}$ and $f_{max}$ denote the global minimum and global maximum values of a function sampled on a manifold, $\Mgroup$. The interval $[f_{min}, f_{max}]$ is partitioned into $L$ equally spaced isolevels. Then, a series of level sets for the isolevels is extracted using the marching triangles algorithm~\cite{hilton1996marching}, and they are overlaid on top of the model.

\paragraph{Interpretation} When observing function topology with isocontours, the most important features are the visual patterns that arise as the contours wrap around the model. Formation or merging of contours will occur around the critical points, so identifying those features may create a better picture of the function's topology. The topological evolution of isocontours is illustrated with red-dotted boxes in \autoref{fig:vis_examples:iso}. 

\paragraph{Interaction} We define the level of detail for isocontours as the number of isolevels, i.e., $L$, where $L$ is a user-selectable parameter. Larger values of $L$ result in higher level of detail, and smaller values of $L$ result in lower level of detail (see the skull dataset example in \autoref{fig:vis_examples:iso}). The parameter $L$ can be modified to change the level of detail, thereby enhancing the perception of the structure of the function. The view can also be rotated and zoomed to overcome occlusion and enhance details.

\subsection{Topology-Based Visualization: Reeb Graphs}
\label{sec:visualization:reeb}
Reeb graphs (see \autoref{fig:vis_examples:reeb}) are used to visualize the topological structure as a skeleton (see \autoref{sec:reebGraphs}). 

\paragraph{Design} Our visualization embeds the Reeb graph onto the model, similar to how it is handled in the Topology Toolkit~\cite{ttk,ttk_overview}. The nodes of the Reeb graph, which represent the critical points of the function, are overlaid on the model. 
The edges represent the flow of the function (contours) between critical points. Therefore, a naive edge connection scheme would not suffice because it would not represent the flow of data correctly. We utilized the centroids of isocontours to draw the trajectory of the arcs throughout the model. 

\paragraph{Interpretation} Since the Reeb graph is embedded in the model, its features can be directly correlated with the function on the model. The primary features to look for in a Reeb graph are the positions of critical points, which are represented as blue dots on the model, and the arc trajectories between blue dots, which are represented as red curves. 

\paragraph{Interaction} We define the level of detail in the context of the Reeb graph as the number of arcs of the graph. For high-density noise, the Reeb graph visualization might look cluttered owing to the high level of detail arising from noise, as shown in the bottom right image of the skull in \autoref{fig:vis_examples:reeb}. Visual clutter can be interactively mitigated by adjusting the level of detail using persistence-guided pruning~\cite{TA:Edelsbrunner:2002:persistenceSimplification}, which removes low persistence critical point pairs from the visualization, as shown in the bottom left image of the skull in \autoref{fig:vis_examples:reeb}. Notably, persistence-based noise reduction may remove longer edges before short ones, which may be counter-intuitive. Occlusion and clutter can be further mitigated by rotating or zooming the view.

\subsection{Topology-Based Visualization: Persistence Diagrams}
\label{sec:visualization:persistence}
Persistence diagrams (see \autoref{fig:vis_examples:pd}) are scatterplots that can be used to illustrate the birth-death pairs of topological data for contour trees, Reeb graphs, and other TDA-based techniques. A birth-death pair corresponds to the pair of minimum/maximum and saddle critical points that result in the creation and disappearance of a connected component, respectively, as described in \autoref{sec:pd}.
Persistence diagrams provide a stable view of the function from a topological perspective, as they show the relationship of the critical points in the Reeb graph.

\paragraph{Design} To generate the persistence diagrams, Reeb graphs are generated for the dataset, and the birth-death pairs are then identified, as described in \autoref{sec:pd}. The pairs are then visualized in the persistence diagram. The horizontal axis of a persistence diagram is labeled as \textit{birth} and denotes the appearance of a feature.
The vertical axis is labeled as \textit{death} and denotes the disappearance of a feature. The blue dot represents local saddle-minimum or saddle-maximum pairs. The red dots represent saddle-saddle pairs, which are formed by holes/tunnels in the model.
The diagonal line indicates where \textit{birth}~$=$~\textit{death}, and the vertical distance from the diagonal is a quantified measure of persistence.
Points near the diagonal have a low persistence (i.e., regarded as noise), whereas those farther away are more important. High- and low-persistence features for the rabbit are highlighted in \autoref{fig:vis_examples:pd}. 

\paragraph{Interpretation} When evaluating data with the persistent diagram, one must note the points far away from the diagonal and their color/type, as those points denote the more persistent features. Secondarily, one may consider the distribution and patterns of points along the diagonal (e.g., feature clustering for the rabbit in \autoref{fig:vis_examples:pd}). 

\paragraph{Interaction} We define the level of detail in the context of persistence diagrams as the number of points representing critical point pairs. We provide the functionality to interactively set a persistence threshold, which removes low-persistence points for observing different levels of detail, which is depicted in the bottom of \autoref{fig:vis_examples:pd}.

%% file: fig.vis.evaluated.tex
\begin{figure*}[!t]
    \centering 
    \subfloat[Color map\label{fig:vis_examples:color}]{\includegraphics[width=0.215\linewidth]{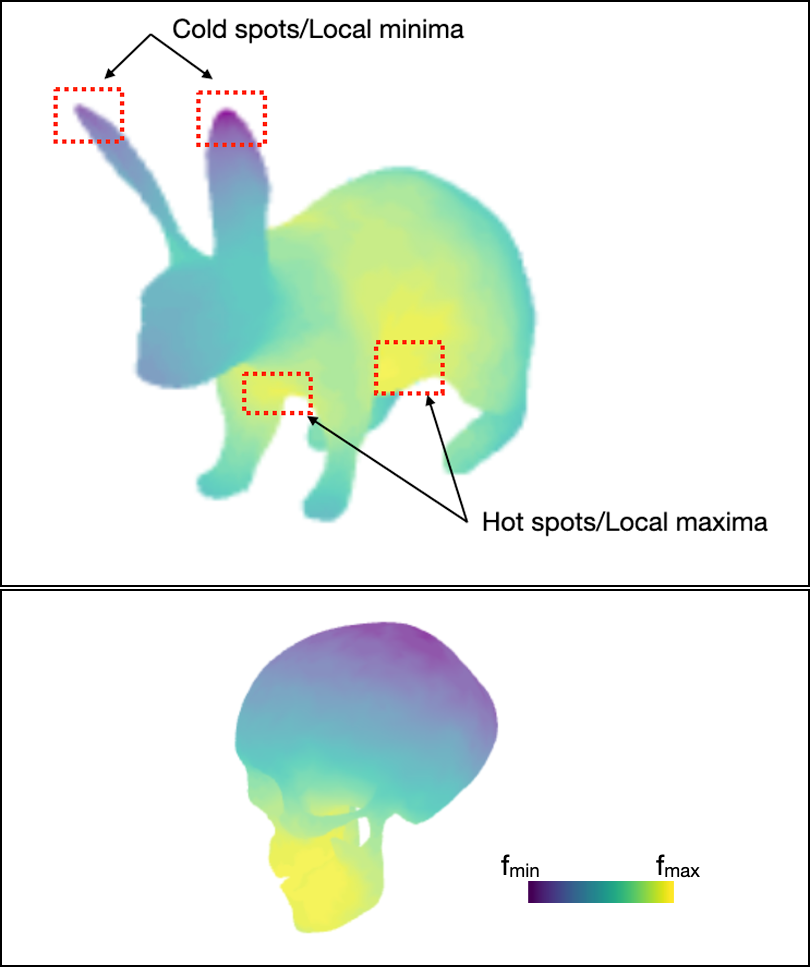}}
    \hspace{7pt}
    \subfloat[Isocontours\label{fig:vis_examples:iso}]{\includegraphics[width=0.215\linewidth]{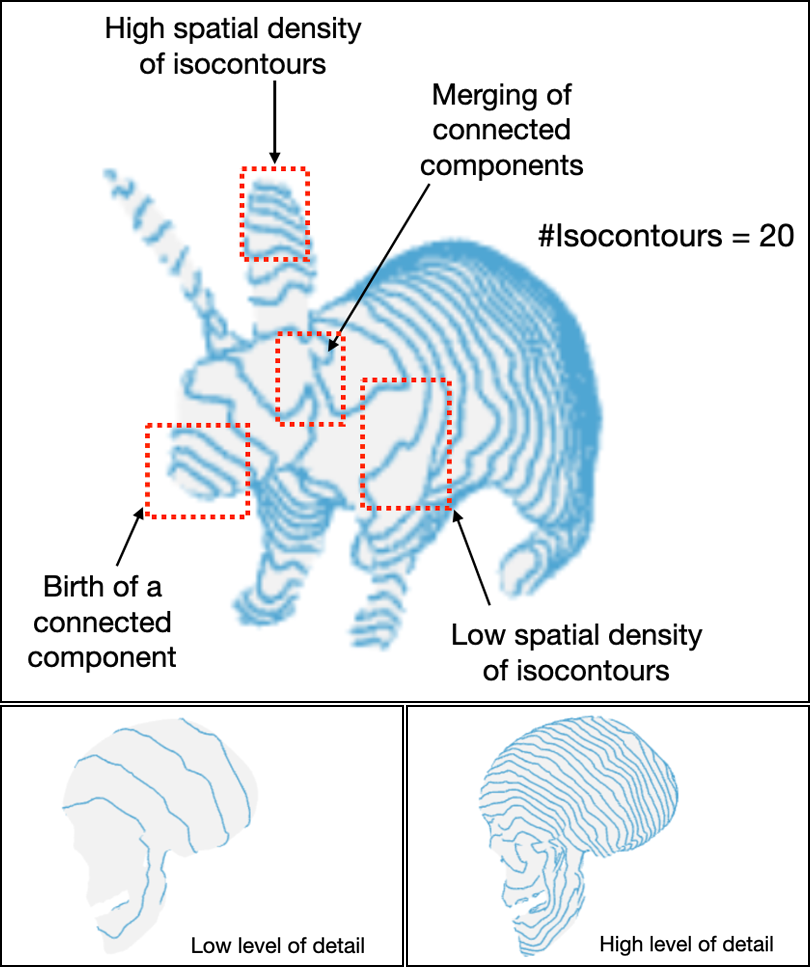}}
    \hspace{7pt}
    \subfloat[Reeb graph\label{fig:vis_examples:reeb}]{\includegraphics[width=0.215\linewidth]{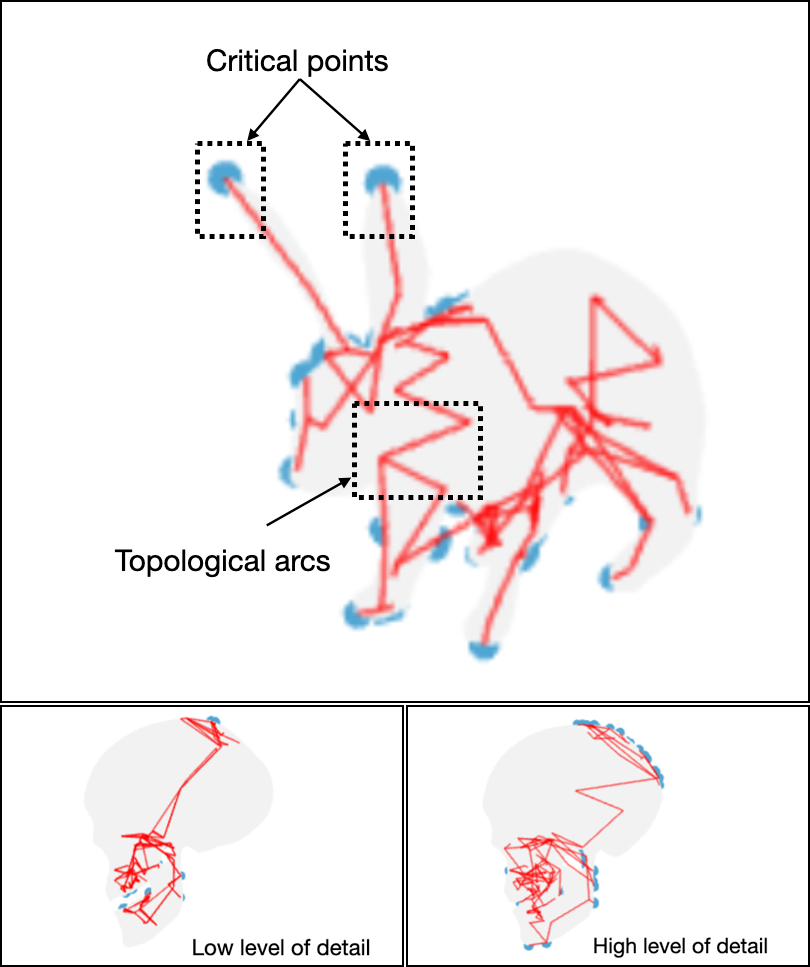}}
    \hspace{7pt}
    \subfloat[Persistence diagram\label{fig:vis_examples:pd}]{\includegraphics[width=0.215\linewidth]{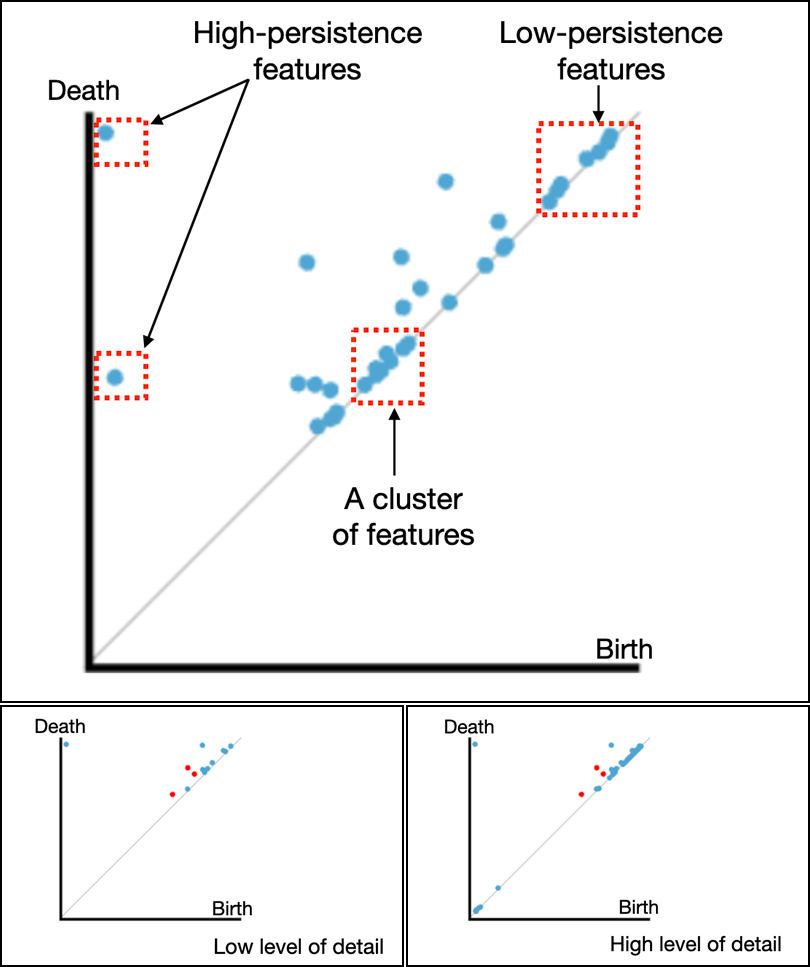}}
    
    \vspace{-7pt}
    \caption{Example visualizations of the rabbit and skull models. (a)~The color map visualization use viridis~\cite{viridis_colormap}, which has a purple hue at the global minimum and a yellow hue at the global maximum. (b)~Important isocontours features are highlighted on the rabbit, while the skull is visualized at low and high levels of detail. (c)~The Reeb graph shows blue spheres at critical point positions and red arcs that indicate the evolution of contours between critical points. The skull is rendered with and without pruning, guided by the persistence of critical points. (d)~In the persistence diagram, the blue dots denote the saddle-minimum or saddle-maximum pairs, and red dots denote the saddle-saddle pairs. The distance from the diagonal is a quantified measure of persistence. Persistence diagrams can be simplified by removing low-persistence features, which are closer to the diagonal.}
    \label{fig:vis_examples}
    \vspace{-5pt}
    
\end{figure*}

%% file: sec-method.tex
\section{Sensitivity Analysis of Topological Visualizations}

We describe the method used for measuring the sensitivity of visualizations to variations in data using a crowdsourced experiment.

\subsection{Data Model}\label{sec:mathematicalModel}

For our study, we model prominent data features as a mixture of 3D Gaussian distributions mapped onto a 2-manifold. Specifically, a given dataset has two components. The first is a 2-manifold triangle mesh, $\Mgroup$, embedded in 3D space. The second component is a scalar field, $f$, which is a mixture of Gaussian distributions and noise defined on the vertices of the mesh. Gaussian mixtures are commonly utilized in scientific literature to model data (e.g., in Vidal et al.~\cite{TA:2020:Vidal:persistenceDiagramBarycenter} and Yan et al.~\cite{TA:2020:Yan:mergeTreeAverage}), and scalar fields on 2-manifolds are observed in many real-world applications (see \autoref{fig:topoVisComparisonRealData}).

\subsubsection{Scalar Field Generation}\label{sec:functionGeneration}

Scalar values were generated per vertex using a combination of several data features and noise. The value at each vertex was calculated as

\vspace{-6pt}
$$f(x,y,z) = \mathcal{N}(x,y,z) + \sum_i^{NOF} \mathcal{G}_i(x,y,z),$$

\vspace{-6pt}
\noindent
where $\mathcal{N}$ was the noise function, $NOF$ (number of features) was the number of salient data features, and $\mathcal{G}_i$ was the data feature function.

\para{Data Features} To simplify the definition of data features, isotropic 3D Gaussian functions were used:

\vspace{-15pt}
$${\displaystyle \mathcal{G}_i(x,y,z)=a_i \cdot \exp \left(-\left(\frac {(x-x_{i})^{2}+(y-y_{i})^{2}+(z-z_{i})^{2}}{2\sigma^{2}}\right)\right)},$$

\vspace{-5pt}
\noindent
where $a_i$ was the amplitude of the feature, which was $1$ by default; $(x_i, y_i, z_i)$ was the source position, which was set using a random vertex on the model; and $\sigma$ was the standard deviation, which was fixed to $1$.

\para{Noise}
Noise was added using the following function:

\vspace{-5pt}
$$\mathcal{N}(x,y,z) = S_N \cdot Perlin(x,y,z),$$

\vspace{-3pt}
\noindent
where $S_N$ is the amplitude of the noise, and $Perlin$ is the noise function. The level of noise was specified using an input signal-to-noise ratio (SNR). Because the default feature amplitude was 1, $S_N=1/SNR$. $Perlin$ was a standard Perlin noise function~\cite{perlin1985image}, the frequency of which was set during our data calibration (see \autoref{sec:calibration}).

\subsection{Sensitivity for 1D Gaussian distributions}

Without a loss of generality, we describe visualization sensitivity in the context of 1D Gaussian distributions. At the end of the section, we discuss how our observations extend to the 3D Gaussian distributions described in the previous section. 

\subsubsection{Position and Amplitude Variation}
\begin{figure}[!hb]
    \centering
    
    \subfloat[Position-based variation\label{fig:exp_variations:sub2}]{\fbox{\includegraphics[width=0.35\linewidth]{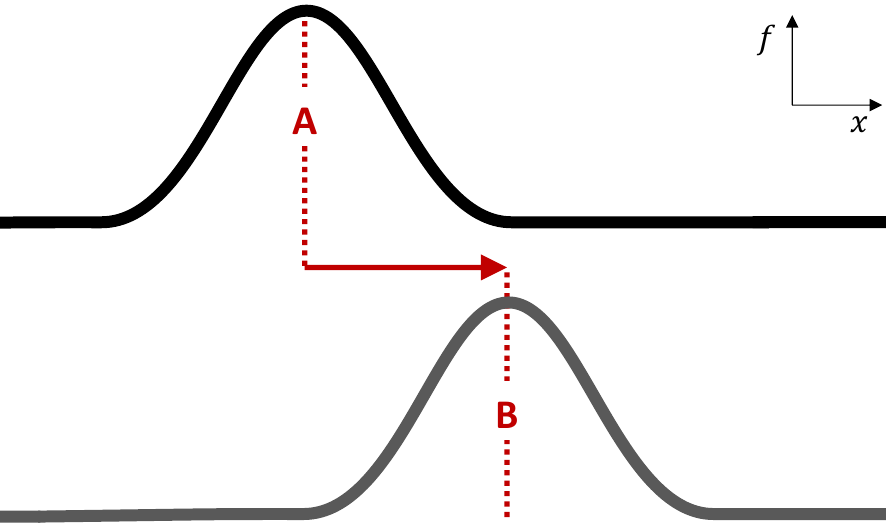}}}
    \hspace{10pt}
    \subfloat[Amplitude-based variation\label{fig:exp_variations:sub1}]{\fbox{\includegraphics[width=0.35\linewidth]{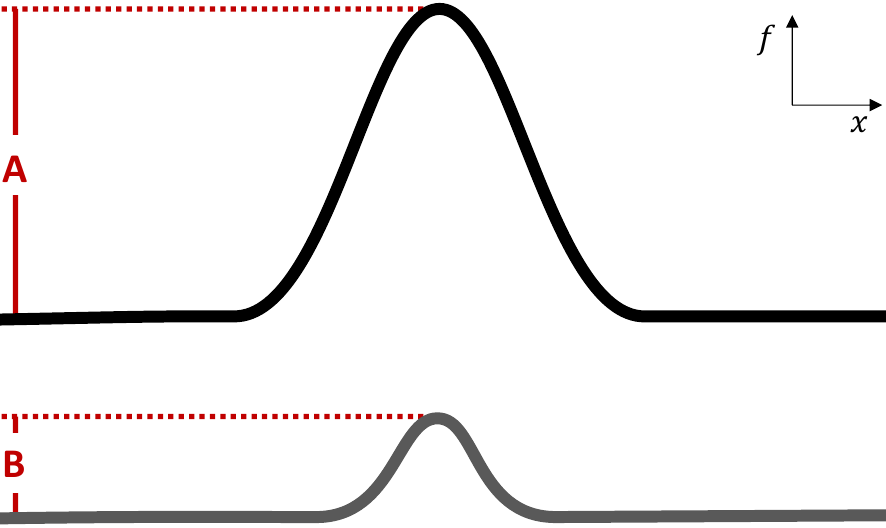}}}
    
    \vspace{-5pt}
    \caption{1D examples of the feature variations studied, including (a)~position variation where the peak at $A$ moves right to $B$, and (b)~amplitude variation where the amplitude of the peak $A$ is reduced in $B$.}
    \label{fig:exp_variations}
    \vspace{-5pt}
\end{figure}

We are interested in evaluating the sensitivity of the topological visualizations described in \autoref{sec:visualization:iso}-\autoref{sec:visualization:persistence} as compared to the reference color mapping (\autoref{sec:visualization:colormap}) to changes in the input function. In particular, we evaluate the sensitivity of these visualization techniques to two variation types. First, we are interested in changes in the \textit{position} of topologically important features of the data. For example, in \autoref{fig:exp_variations:sub2}, the feature of interest is moved from $A$ to $B$. Second, we are interested in sensitivity to changes in the scale or \textit{amplitude} of topologically important features. As shown in \autoref{fig:exp_variations:sub1}, the maximum value of the peak decreases from $A$ to $B$.

\input{sec-method-sensitivity}

\subsection{Generating Positional and Amplitude Variations in 3D}\label{sec:posAmpVariation}

To test the sensitivity, three scalar functions are generated for each experimental trial by first selecting the parameters, as described in \autoref{sec:exp:data}. Initially, a baseline dataset DB is generated. The two additional test datasets are generated by first selecting one feature at random to vary by position or amplitude based on the test being performed. Then, for positional variations, the associated feature Gaussian is moved to $A_0$ or $A_1$ by selecting (at random) a vertex located at a target geodesic distance away from the baseline location. We compute the geodesic distance on the mesh manifold by using Dijkstra's algorithm on the vertices of the mesh, where edge weights are the Euclidean distance between vertices. For amplitude variations, the selected feature had its amplitude modified to the target amplitude of $A_0$ or $A_1$. All other features remained unchanged.

%% file: sec-method-sensitivity.tex
\subsubsection{Measuring Sensitivity} \label{sec:measureSensitivity}

To test the sensitivity of the visualizations, we consider a scenario in which a participant must pick from two experimental visualizations which is the most similar to a baseline visualization. Let $\text{DB}$ denote the baseline dataset corresponding to the original position and amplitude parameters $B$.
Let $\text{DA}_0$ and $\text{DA}_1$ denote the two datasets corresponding to parameters $A_0$ and $A_1$, respectively, representing either variations of amplitude or position for a single feature.
Let $\text{VA}_{0}$, $\text{VA}_{1}$, and $\text{VB}$ be visualizations generated using the \textit{same} visualization technique, either color maps, isocontours, Reeb graphs, or persistence diagrams. We present a participant with these visualizations, and the participant has to decide which of $\text{VA}_{0}$ and $\text{VA}_{1}$ is closer to $\text{VB}$.

\begin{figure}[!t]
    \centering %
    
    \rotatebox{90}{\footnotesize\hspace{10pt}$d(A_0,B)\approx d(A_1,B)$}
    \fbox{\includegraphics[height=2.5cm]{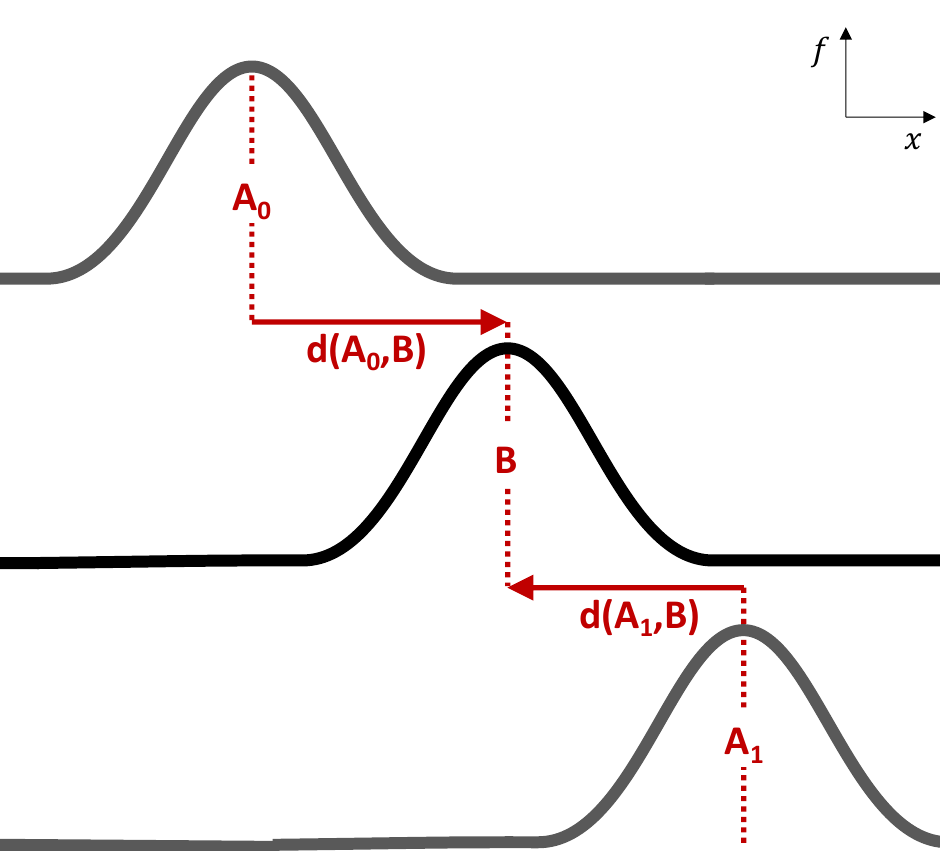}}
    \hspace{12pt}
    \fbox{\includegraphics[height=2.5cm]{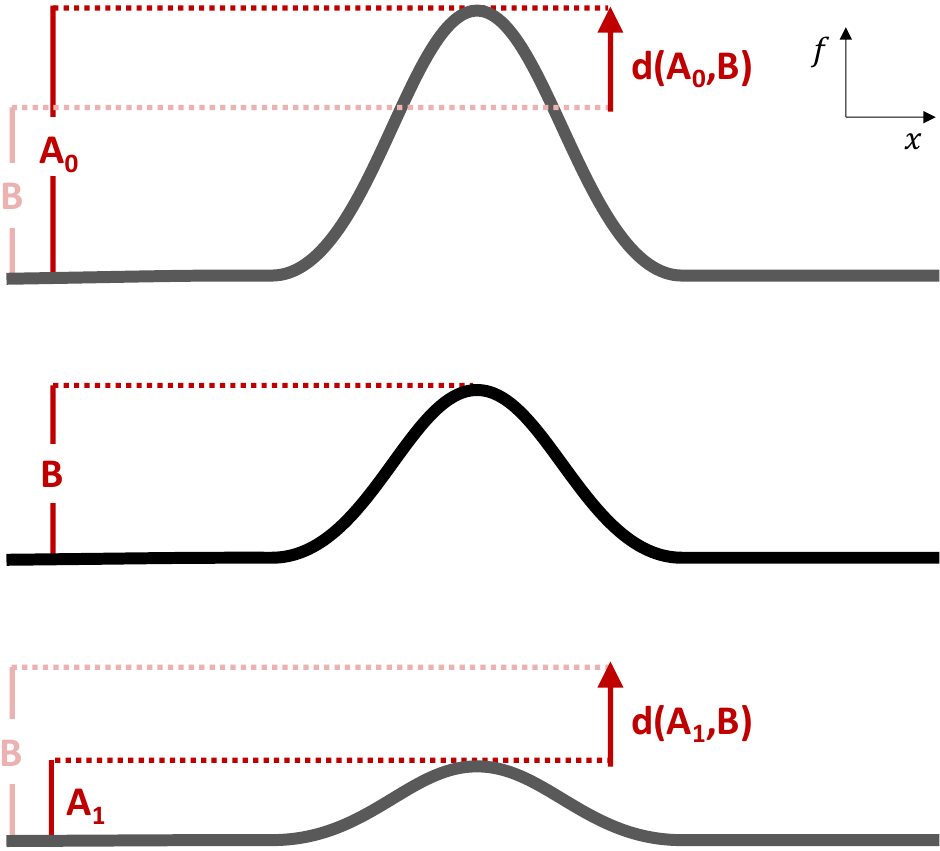}}
    
    \rotatebox{90}{\footnotesize\hspace{10pt}$d(A_0,B) > d(A_1,B)$}
    \subfloat[Positional Variation\label{fig:exp_measure_variations:pos}]{\fbox{\includegraphics[height=2.5cm]{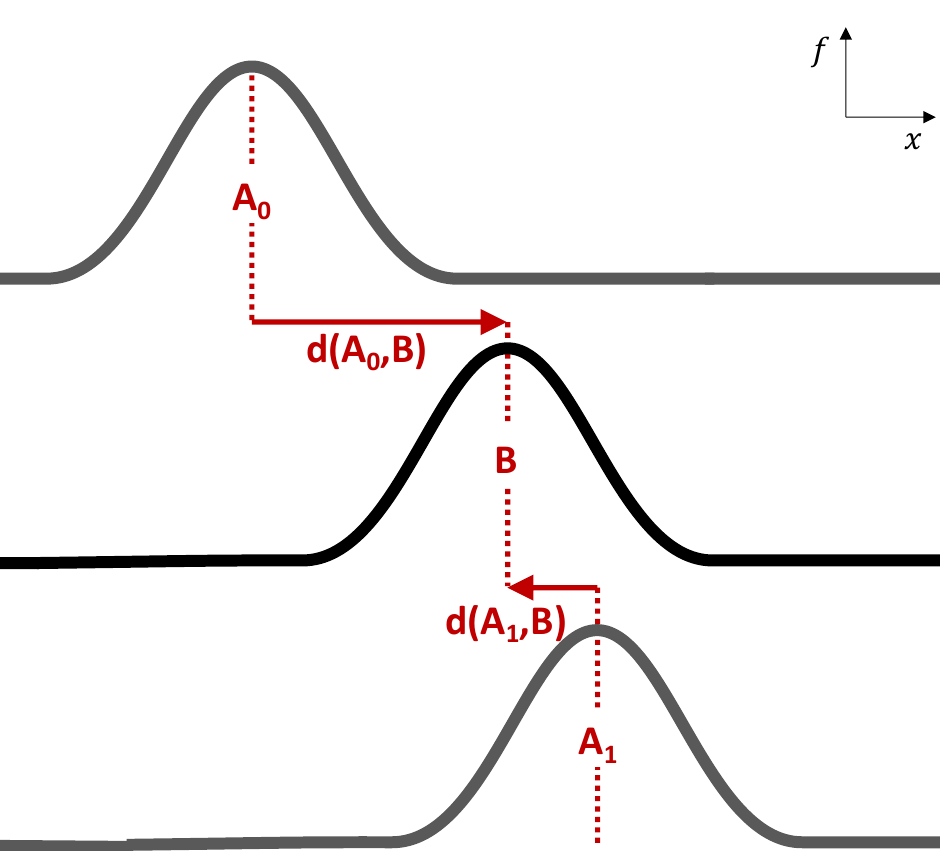}}}
    \hspace{14pt}
    \subfloat[Amplitude Variation\label{fig:exp_measure_variations:amp}]{\fbox{\includegraphics[height=2.5cm]{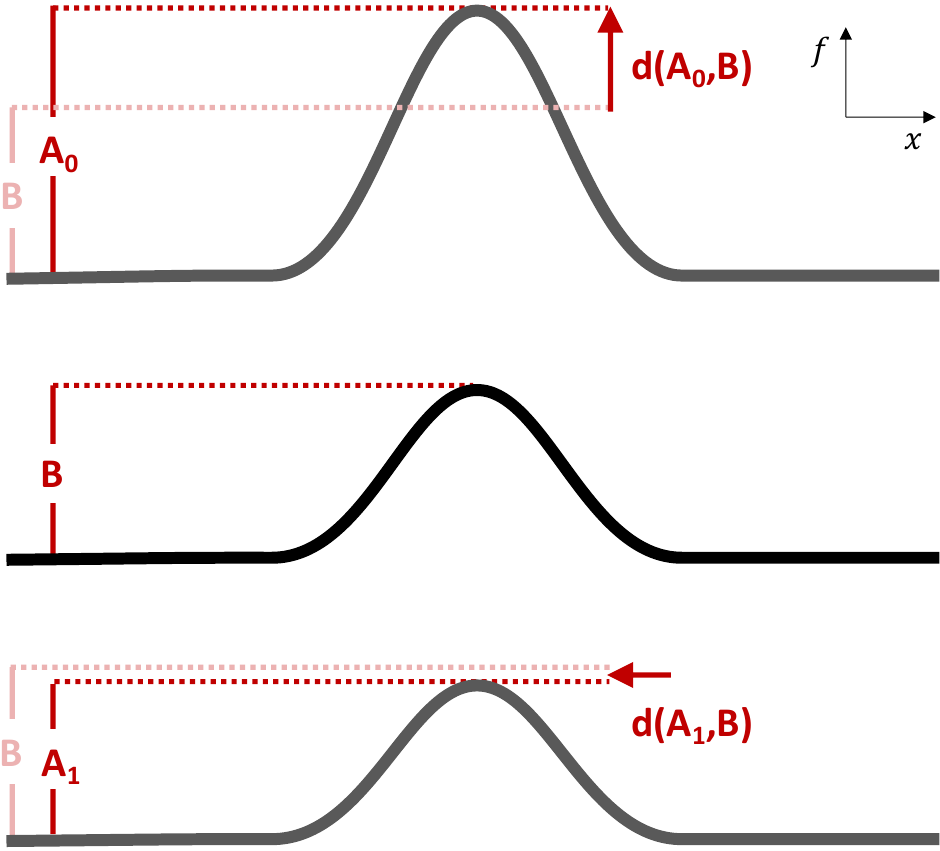}}}
    
    \vspace{-5pt}
    \caption{The illustration of 1D visualization sensitivity to (a)~positional and (b)~amplitude variations. For variation types where $d(A_0,B) \approx d(A_1,B)$ (top), it is difficult to visually determine if the 1D plots for $A_0$ or $A_1$ is closer to the 1D plot for $B$. In contrast, when $d(A_0,B)$ is considerably greater than $d(A_1,B)$ (bottom), it is easier to determine that the plot for parameter $A_1$ is closer to the baseline than the plot for $A_0$.}
    \label{fig:exp_measure_variations}
    \vspace{-5pt}
    
\end{figure}

Consider the 1D examples in \autoref{fig:exp_measure_variations}. For positional variation (see \autoref{fig:exp_measure_variations:pos}), given a baseline visualization for the position parameter $B$, the participant must select between the visualizations for parameters $A_0$ or $A_1$, which they believe is more similar to the baseline visualization. Similarly, for the amplitude variation (see \autoref{fig:exp_measure_variations:amp}), the participant must select between the visualizations for parameters $A_0$ or $A_1$, which is more similar to the baseline visualization. For amplitude variation, either $A_0$ or $A_1$ is always larger than $B$, and the other is smaller than $B$ to ensure the participants are not directly comparing $A_0$ and~$A_1$.

To evaluate the sensitivity of a visualization to variations in features, we measure how often participants can correctly select between the $\text{VA}_{0}$ and $\text{VA}_{1}$ that is closer to $\text{VB}$ when considering the following distance measure between the dataset parameters: 

\noindent
$$A'=\big|d(A_0,B)-d(A_1,B)\big|,$$

\vspace{-4pt}
\noindent
where $d$ indicates the distance between the positional or amplitude parameters. For positional variations, $d(A,B)$ is the geodesic distance between the location of the feature being moved in $A$ and $B$. For amplitude variations, the absolute difference between the amplitude of the feature being modified is used (i.e., $d(A,B) = |A-B|$). 

Intuitively, $A'$ measures how different in distance or amplitude the variations are from the baseline. A larger $A'$ value implies that one stimulus is much more similar to the baseline and would therefore be more likely to be selected by a participant. Consider the example shown in \autoref{fig:exp_measure_variations}. For both positional and amplitude variations, it is not easy to decide which visualization is perceptually more similar to the baseline visualization when $d(A_0,B) \approx d(A_1,B)$ (see \autoref{fig:exp_measure_variations}(top)). In contrast, when $d(A_1,B)$ is significantly smaller than $d(A_0,B)$ (see \autoref{fig:exp_measure_variations}(bottom)), it is relatively easy to visually decide that the 1D plot for parameter $A_1$ is closer to the baseline visualization.

Generally speaking, the sensitivity of a quantity of interest Y with respect to a parameter X is defined as the first derivative $\partial Y$/$\partial X$ and can be estimated through linear regression for the observed data~\cite{saltelli2008global}. In perceptual psychology, however, the Weber-Fechner Law~\cite{fechner1948elements} notes that the minimal detectable increase in the stimulus is proportional to the stimulus magnitude (Weber's law), and the perceived intensity is proportional to logarithm of the actual intensity (Frechner's inference). Thus, the sensitivity of visualizations is captured by the change in the accuracy of user decisions with respect to the change in parameter $A'$ using logistic regression (see the analysis description in \autoref{sec:data_analysis}).

%% file: sec-exp_env.tex
\section{Experiment}

\subsection{Hypotheses}

To understand the sensitivity of the four visualizations, we consider the topological aspects of the function that each visualization shows. \autoref{fig:teaser} illustrates the sensitivity of each visualization to positional and amplitude variations, assisting in the formulation of our hypotheses.

\para{Color Maps} For color mapping, we have chosen the viridis color map (see \autoref{sec:visualization:colormap} for more details). Thus, the brightest spot in the color map should also indicate the location of the peak of the feature. Therefore, we hypothesize that \textit{color maps will be sensitive to variations in both the amplitude and the position of features.}

\para{Isocontours} Similar to color maps, the concentric rings of isocontours show where function extrema occur. However, the rings themselves provide no direct indication of the function value. Therefore, we hypothesize that \textit{isocontours will be sensitive to variations in the position of features but not sensitive to the amplitude of features.}

\para{Reeb Graphs} Reeb graphs show the interconnectedness of critical points in the function. However, they do not provide any indication of the value of those critical points. Therefore, we hypothesize that \textit{Reeb graphs will be sensitive to variations in the position of features but not sensitive to the amplitude of features.}

\para{Persistence Diagrams} Persistence diagrams show only the birth and death of critical features, which correspond to the amplitude of a birth-death pair. Positional variations are visible indirectly only when pairs switch. Therefore, we hypothesize that \textit{persistence diagrams will be sensitive to variations in the amplitude of features but not sensitive to their position.}

\subsection{Experimental Interface}

The experiment consisted of a web page with four demographic questions, a tutorial describing all of the visualization types and interactive capabilities (e.g., rotation and zooming), four practice questions, 24 experimental trials, and a post-experiment questionnaire. The tutorial and practice questions serve to familiarize participants, especially those with no expertise or limited prior experience in visualization, with the four visualization types. Specifically, we present guidelines for participants summarizing what specific patterns to look for when decoding the visualizations (similar to the interpretation descriptions in \autoref{sec:visualization}).
Please refer to pages 4-10 of the supplementary material illustrating the tutorial and practice questions for a single user. The responses to our post-experiment questionnaire (see pp.\ 39-47 of the supplementary material and~\autoref{sec:qualitativeFeedback}) capture the thought process of participants and the difficulties they encountered when interpreting the four visualization types and making decisions.

\begin{figure}[!b]
    \centering

    \vspace{-7pt}
    \includegraphics[width=0.75\linewidth]{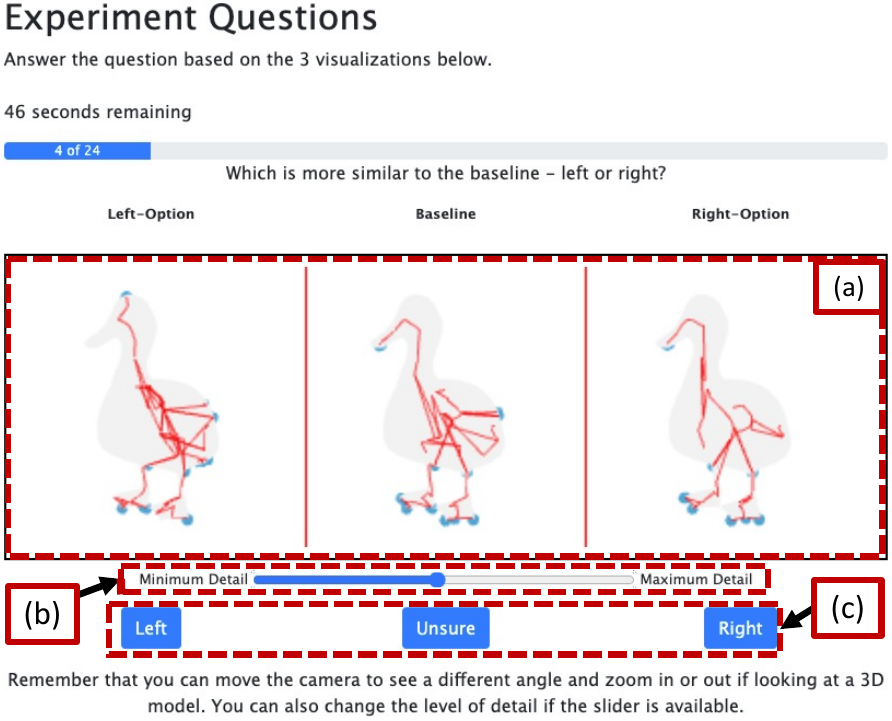}
    
    \vspace{-5pt}
    \caption{Example of the experimental interface on Reeb graphs. (a)~The three visualizations include a baseline in the center and experimental choices to the left and right. The navigation was coordinated such that changes in orientation or zoom were applied to all three visualizations. (b)~Visualizations that allow level-of-detail adjustments had a slider for this purpose. (c)~The participant had to make one of three choices.}    
    \label{fig:exp_ex}
    
\end{figure}

The main experimental interface (see \autoref{fig:exp_ex}) consists of the three visualizations, always of the same type, including a baseline in the center and experimental dataset choices to the left and right (see \autoref{fig:exp_ex}(a)). An orbiting camera with zoom was used so participants could view the models from every angle, which enabled the participants to observe features that would otherwise be hidden and to investigate details. The navigation was coordinated to ensure that changes to orientation or zoom were applied to all three visualizations.
Visualizations that allow level-of-detail adjustments had a slider for this purpose (see \autoref{fig:exp_ex}(b)). 
We asked the participants which of the two experimental visualization choices was more similar to the baseline (see \autoref{fig:exp_ex}(c)). 
The participants had to make one of three choices: (1)~left, (2)~right, or (3) unsure if they were unable to answer with any confidence. 
The correct answer (i.e., the more similar visualization) was randomly shuffled between the left and right sides. For each trial, the participant was allowed up to 60 seconds, after which the experiment automatically advanced.

\subsection{Model Selection}

Models were selected to balance the desire for realistically complex geometry while also limiting cognitive strain caused by interpreting the model. We, therefore, selected familiar 3D shapes, namely, biological models from humans and animals (see examples in the supplemental material). To avoid categorical bias, we selected four models from each of the categories: human busts, human anatomical models, human extremities, land animals, sea animals, and birds.
The goldenRetriever, lion, rabbit, horse, skull, tooth, turtle, shark, fish, owl, parrot, and bird come from the Princeton Shape Benchmark repository~\cite{princeton_shape_benchmark}. The bimba, bust, windfish, handFist, and handPointPrep are from the AIM@SHAPE repository~\cite{digitalShapeWorkbench}.  The heart and kidney are from the University of Michigan's BlueLink AnatomyTOOL~\cite{bluelinkanatomy}. The duck, frederick, lincoln, foot1, and foot2 were acquired from Free3D~\cite{printablemodels}. 
The models were prepared and cleaned using Blender 3D's~\cite{blender} mesh retopology and malformed mesh detection tools. All models were then normalized to fit in a unit box at the origin.

%% file: sec-experiment.tex
\subsection{Variables}

The independent variables of our experiment include the 3D model; visualization types (color maps, isocontours, Reeb graphs, or persistence diagrams); variation type (position or amplitude); values of $A_0$, $A_1$, and $A'$; SNR; and the NOF. However, we specifically focus on evaluating visualization types, variation types, and $A'$ values.
The dependent variables are the \textit{accuracy} and \textit{sensitivity} of selection, time taken, mouse interactions (rotate and zoom), and level-of-detail interactions.

\subsection{Parameter Calibration}
\label{sec:calibration}

Determining parameter ranges was an important problem because the data complexity had to be tuned to ensure that the task was neither too difficult nor too easy for all visualization types. In particular, we were focused on identifying reasonable values of $A'$, SNR, and NOF.
Our parameters went through a four-stage calibration process. (1)~Initially, parameters were set using the research team's observations of values. (2)~Next, we ran an internal study using the research team and lab members to check the difficulty of tasks and adjust accordingly. (3)~Then, we ran a small-scale (40-person) preliminary study on Amazon Mechanical Turk. (4)~A final internal study was repeated using the research team and lab members. At each stage, parameters were adjusted to calibrate difficulty and experiment length and to improve the experimental interface.

\subsection{Data Generation}
\label{sec:exp:data}

A Python script was used to generate the model and function configuration for all participants in the experiment. The within-subject design consisted of 24 trials. Each participant was presented with trials that had the following characteristics:
\begin{itemize}[noitemsep]
    \item 1 trial per 3D model
    \item 6 trial per visualizations type: \{color map, isocontour, Reeb graph, persistence diagram\}
    \item 12 trial per variation type: \{position, scale\}
    \item 6 trial per $A'$: $\{0.15, 0.30, 0.45, 0.60\}$
    \item 8 trial per SNR value: $\{80, 90, 100\}$
    \item 3 trial per NOF in the range: $[2, 9]$
\end{itemize}
Thus, we ensured that each user was presented with the datasets generated using a balanced distribution of parameters, including visualization type, function type, $A'$, SNR, and NOF. Additionally, $A_0$ was randomly selected such that $|A_0|\in[0.1, 0.9]$, and $A_1$ was selected such that $|A_1|\in[0.1, 0.9]$ while also satisfying the $A'$ requirement. For amplitude variation, $A_0$ is added to $B$, and $A_1$ is subtracted from $B$, or vice versa, in a random manner to guarantee that one stimulus would have a larger amplitude, and one stimulus would have a smaller amplitude relative to the baseline.

\subsection{Data Collection and Analysis Methodology}
\label{sec:data_analysis}

Datasets were pre-generated using a Python script and the parameters described \autoref{sec:exp:data}. The model, function, Reeb graph, and persistence data were stored in JSON format and ZIP compressed to improve the file transfer rate. The experiment was run on a custom-built Node.js web server. 
Three.js~\cite{threejs} was used for color map, isocontour, and Reeb graph rendering. D3.js~\cite{bostock2011d3} was used for persistent diagram rendering. The visualizations were custom implementations.
The answers from each participant were recorded in a JSON document with entries containing participant ID, visualization type, parameters for the stimuli, participant's selection, interaction information (i.e., click, scroll, slider move counts), and time taken.

We utilized three statistical tools in the analysis. 
First, a binomial test, which determines if an observed distribution deviated from an expected distribution, was used to determine whether the accuracy of the visualizations was statistically significant from a null hypothesis of 50\% (i.e., guessing would achieve a score of approximately 50\%). 
Next, to determine if the differences in accuracy for each method were significant, we utilized a $\chi^2$ contingency test, which determines if the observed distributions in one or more categories deviated from an expected distribution with a null hypothesis that all methods had identical accuracy.
Finally, binary logistical regression is commonly used as a statistical model for hypothesis testing of a continuous independent variable and binary dependent variable~\cite{hosmer2013applied}. We used a logit function (i.e., binary logistical regression) to evaluate whether the accuracy of visualization methods was sensitive to increases in $A'$ with a null hypothesis that it was not. 
For all tests, we consider significance to be $p<.05$, but we report exact $p$ values to 3 digits for completeness.

%% file: sec-results.tex
\section{Results}

We conducted the institutional review board-approved study using participants from Amazon Mechanical Turk. The experiment generally took less than 15 minutes, and each participant was compensated \$2 USD. There were 120 participants filtered by region (US and Canada) and HIT Approval Rate ($>95\%$), of which 18 provided problematic data (i.e., the participant failed to engage with the study because they did not answer the complete 24 questions, frequently timed out during the experiment, or their median response time was less than 5~seconds). Hence, we analyzed 2,448 trials from the remaining 102 participants.

\begin{figure}[!b]
    \centering 
    \vspace{-5pt}
    \subfloat[Age distribution\label{fig:age_pie_chart}]{{\includegraphics[height=2.7cm]{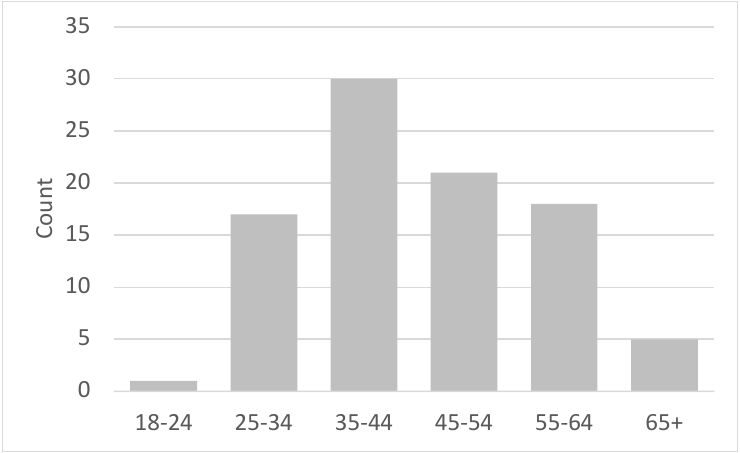}}}
    \hspace{5pt}
    \subfloat[Visualization experience\label{fig:vis_experience_pie_chart}]{{\includegraphics[height=2.7cm]{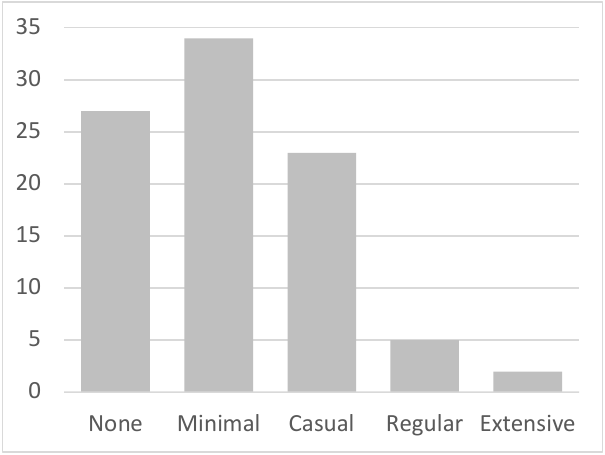}}}
    
    \vspace{-5pt}
    \caption{Bar charts of the demographic survey of participants.}
\end{figure}

\paragraph{Participant Demographics}
57\% of the participants were male, 42\% female, and 1\% nonbinary. All participants were 18 years old or older (see \autoref{fig:age_pie_chart}). 92\% of the participants reported having casual, minimal, or no visualization experience (see \autoref{fig:vis_experience_pie_chart}).

In \autoref{sec:task_accuracy}, we present the overall accuracy of participants for position and amplitude variations in data features.  Next, \autoref{sec:task_sensitivity} reports the sensitivity of participants' decisions to feature variations. Finally, we report observations of the timing and interactions and collected feedback from participants in \autoref{sec:task_time_accuracy} and \autoref{sec:qualitativeFeedback}, respectively.

\renewcommand{\arraystretch}{1.2}

\begin{figure*}[!htb]
    \centering
    
    \rotatebox{90}{\hspace{20pt}\textit{Positional} variation}
    \subfloat[Color maps ($p=.613$)\label{fig:colormap_position_logit}]{
    \begin{minipage}[t]{0.23\linewidth}
    \includegraphics[width=\linewidth]{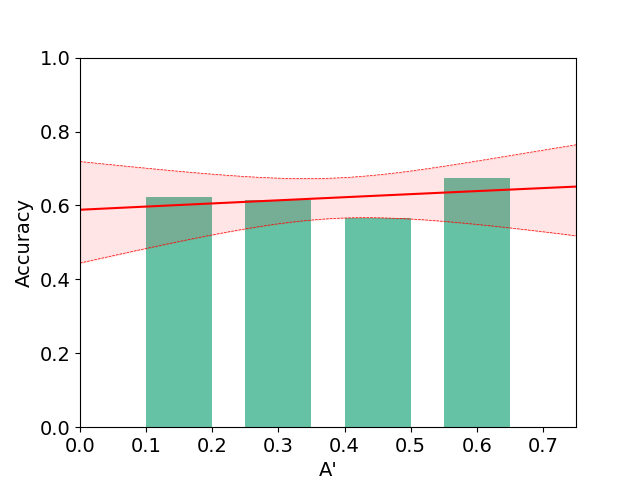}
    \resizebox{\linewidth}{!}{%
    \begin{tabular}{@{\hspace{3pt}}c@{\hspace{3pt}}|@{\hspace{3pt}}c@{\hspace{3pt}}|@{\hspace{3pt}}c@{\hspace{3pt}}|@{\hspace{3pt}}c@{\hspace{3pt}}|@{\hspace{3pt}}c@{\hspace{3pt}}|@{\hspace{3pt}}c@{\hspace{3pt}}}
        Param.   & Coef. & SE & z & $P>|z|$ & 95\% CI \\
        \hline
        $A'$        & $0.355$  & $0.703$ & $0.505$  & $0.613$ & $[$-$1.022, 1.732]$ \\
        (intcp) & $0.357$ & $0.297$ & $1.204$ & $0.228$ & $[$-$0.224, 0.939]$ \\
    \end{tabular}}
    \end{minipage}
    }
    \hfill
    \subfloat[Isocontours ($p=.991$)\label{fig:isocontour_position_logit}]{
    \begin{minipage}[t]{0.23\linewidth}
    \includegraphics[width=\linewidth]{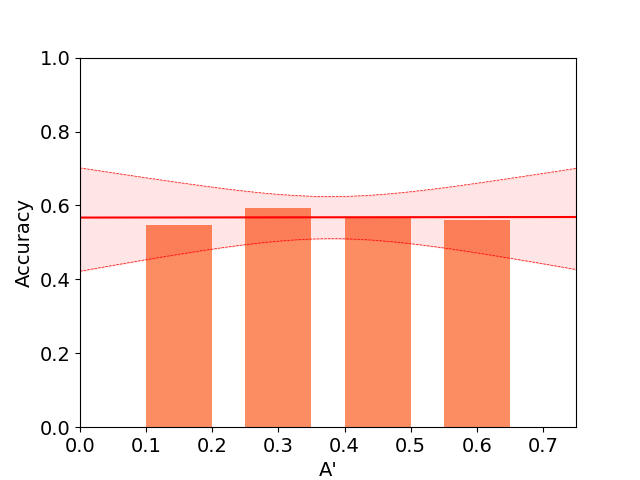}
    \resizebox{\linewidth}{!}{%
    \begin{tabular}{@{\hspace{3pt}}c@{\hspace{3pt}}|@{\hspace{3pt}}c@{\hspace{3pt}}|@{\hspace{3pt}}c@{\hspace{3pt}}|@{\hspace{3pt}}c@{\hspace{3pt}}|@{\hspace{3pt}}c@{\hspace{3pt}}|@{\hspace{3pt}}c@{\hspace{3pt}}}
        Param.   & Coef. & SE & z & $P>|z|$ & 95\% CI \\
        \hline
        $A'$        & $0.008$  & $0.720$ & $0.011$  & $0.991$ & $[$-$1.403, 1.419]$ \\
        (intcp) & $0.270$ & $0.298$ & $0.906$ & $0.365$ & $[$-$0.314, 0.855]$ \\
    \end{tabular}}   
    \end{minipage}
    }
    \hfill
    \subfloat[Reeb graphs ($\bm{p=.034}$)\label{fig:reebgraph_position_logit}]{
    \begin{minipage}[t]{0.23\linewidth}
    \includegraphics[width=\linewidth]{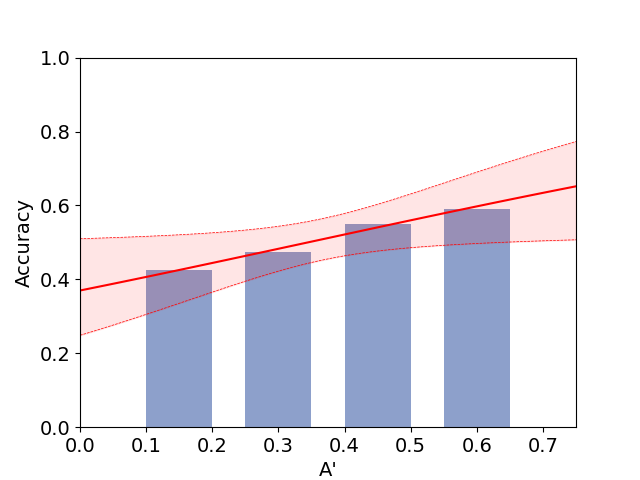}
    \resizebox{\linewidth}{!}{%
    \begin{tabular}{@{\hspace{3pt}}c@{\hspace{3pt}}|@{\hspace{3pt}}c@{\hspace{3pt}}|@{\hspace{3pt}}c@{\hspace{3pt}}|@{\hspace{3pt}}c@{\hspace{3pt}}|@{\hspace{3pt}}c@{\hspace{3pt}}|@{\hspace{3pt}}c@{\hspace{3pt}}}
        Param.   & Coef. & SE & z & $P>|z|$ & 95\% CI \\
        \hline
        $A'$        & $1.546$  & $0.735$ & $2.103$  & $0.035$ & $[0.105, 2.987]$ \\
        (intcp) & -$0.533$ & $0.292$ & -$1.823$ & $0.068$ & $[$-$1.106, 0.040]$ \\
    \end{tabular}}
    \end{minipage}
    }
    \hfill
    \subfloat[Persistence diagrams  ($p=.276$)\label{fig:persistencediagram_position_logit}]{
    \begin{minipage}[t]{0.23\linewidth}
    \includegraphics[width=\linewidth]{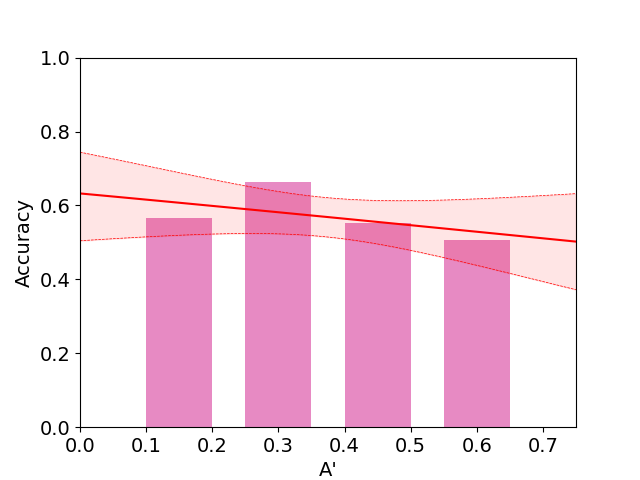}
    \resizebox{\linewidth}{!}{%
    \begin{tabular}{@{\hspace{3pt}}c@{\hspace{3pt}}|@{\hspace{3pt}}c@{\hspace{3pt}}|@{\hspace{3pt}}c@{\hspace{3pt}}|@{\hspace{3pt}}c@{\hspace{3pt}}|@{\hspace{3pt}}c@{\hspace{3pt}}|@{\hspace{3pt}}c@{\hspace{3pt}}}
        Param.   & Coef. & SE & z & $P>|z|$ & 95\% CI \\
        \hline
        $A'$        & -$0.712$  & $0.655$ & -$1.087$  & $0.277$ & $[$-$1.997, 0.572]$ \\
        (intcp) & $0.543$ & $0.268$ & $2.027$ & $0.043$ & $[0.018, 1.068]$ \\
    \end{tabular}}
    \end{minipage}
    }

    \vspace{-6pt}
    \rotatebox{90}{\hspace{20pt}\textit{Amplitude} variation}
    \subfloat[Color maps ($\bm{p=.010}$)\label{fig:colormap_scale_logit}]{
    \begin{minipage}[t]{0.23\linewidth}
    \includegraphics[width=\linewidth]{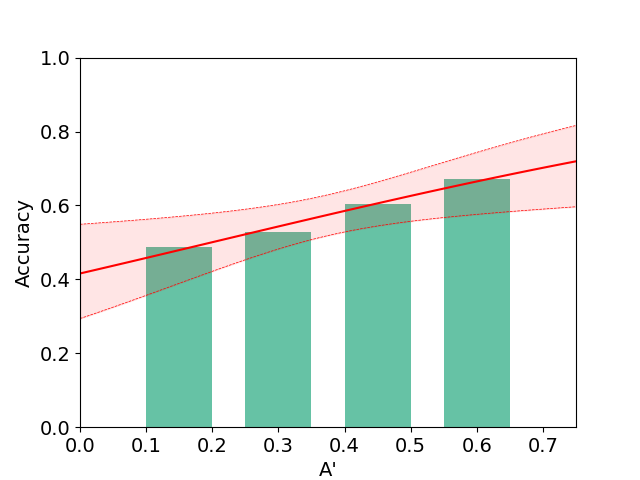}
    \resizebox{\linewidth}{!}{%
    \begin{tabular}{@{\hspace{3pt}}c@{\hspace{3pt}}|@{\hspace{3pt}}c@{\hspace{3pt}}|@{\hspace{3pt}}c@{\hspace{3pt}}|@{\hspace{3pt}}c@{\hspace{3pt}}|@{\hspace{3pt}}c@{\hspace{3pt}}|@{\hspace{3pt}}c@{\hspace{3pt}}}
        Param.   & Coef. & SE & z & $P>|z|$ & 95\% CI \\
        \hline
        $A'$        & $1.708$  & $0.672$ & $2.541$  & $0.011$ & $[0.391, 3.026]$ \\
        (intcp) & -$0.340$ & $0.274$ & -$1.241$ & $0.215$ & $[$-$0.876, 0.197]$ \\
    \end{tabular}}
    \end{minipage}
    }
    \hfill
    \subfloat[Isocontours ($p=.059$)\label{fig:isocontour_scale_logit}]{
    \begin{minipage}[t]{0.23\linewidth}
    \includegraphics[width=\linewidth]{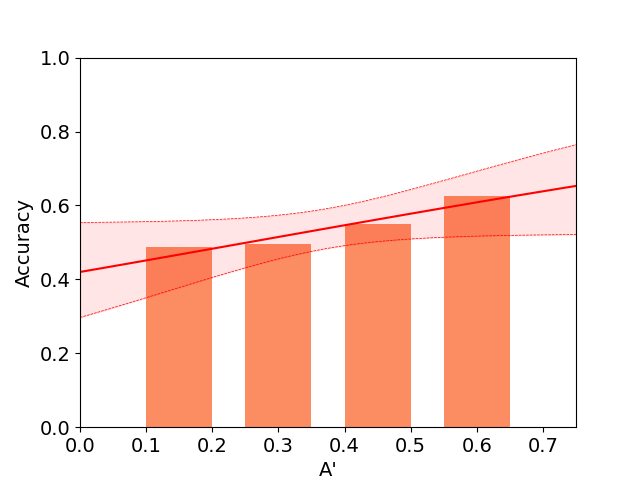}
    \resizebox{\linewidth}{!}{%
    \begin{tabular}{@{\hspace{3pt}}c@{\hspace{3pt}}|@{\hspace{3pt}}c@{\hspace{3pt}}|@{\hspace{3pt}}c@{\hspace{3pt}}|@{\hspace{3pt}}c@{\hspace{3pt}}|@{\hspace{3pt}}c@{\hspace{3pt}}|@{\hspace{3pt}}c@{\hspace{3pt}}}
        Param.   & Coef. & SE & z & $P>|z|$ & 95\% CI \\
        \hline
        $A'$        & $1.272$  & $0.676$ & $1.881$  & $0.060$ & $[$-$0.054, 2.598]$ \\
        (intcp) & -$0.323$ & $0.275$ & -$1.173$ & $0.241$ & $[$-$0.862, 0.217]$ \\
    \end{tabular}}
    \end{minipage}
    }
    \hfill
    \subfloat[Reeb graphs ($p=.817$)\label{fig:reebgraph_scale_logit}]{
    \begin{minipage}[t]{0.23\linewidth}
    \includegraphics[width=\linewidth]{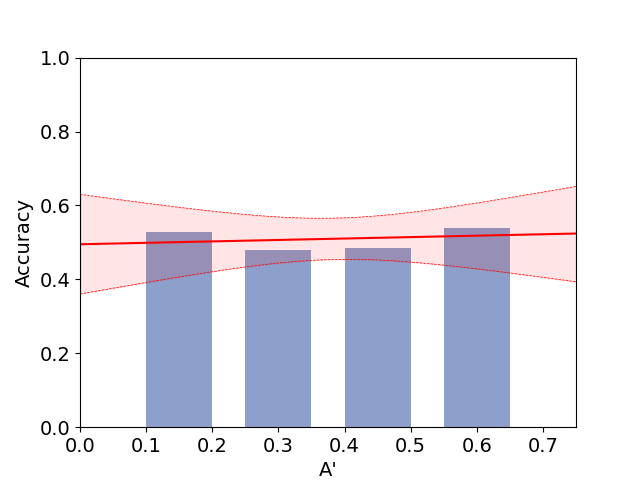}
    \resizebox{\linewidth}{!}{%
    \begin{tabular}{@{\hspace{3pt}}c@{\hspace{3pt}}|@{\hspace{3pt}}c@{\hspace{3pt}}|@{\hspace{3pt}}c@{\hspace{3pt}}|@{\hspace{3pt}}c@{\hspace{3pt}}|@{\hspace{3pt}}c@{\hspace{3pt}}|@{\hspace{3pt}}c@{\hspace{3pt}}}
        Param.   & Coef. & SE & z & $P>|z|$ & 95\% CI \\
        \hline
        $A'$        & $0.155$  & $0.670$ & $0.231$  & $0.817$ & $[$-$1.158, 1.467]$ \\
        (intcp) & -$0.020$ & $0.282$ & -$0.071$ & $0.943$ & $[$-$0.573, 0.533]$ \\
    \end{tabular}}
    \end{minipage}
    }
    \hfill
    \subfloat[Persistence diagrams ($\bm{p<.001}$)\label{fig:persistence_diagram_scale_logit}]{
    \begin{minipage}[t]{0.23\linewidth}
    \includegraphics[width=\linewidth]{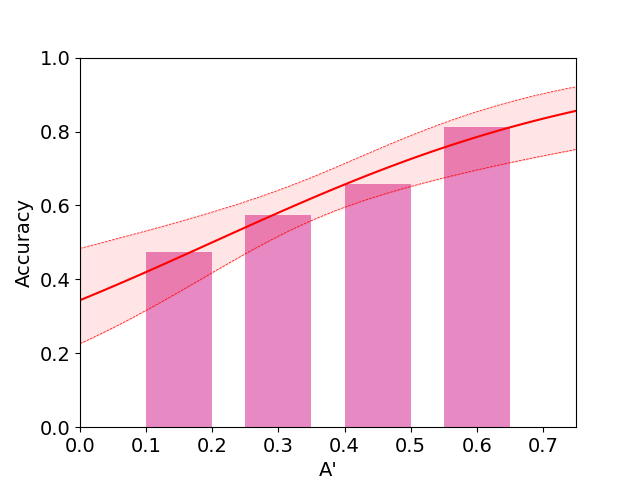}
    \resizebox{\linewidth}{!}{%
    \begin{tabular}{@{\hspace{3pt}}c@{\hspace{3pt}}|@{\hspace{3pt}}c@{\hspace{3pt}}|@{\hspace{3pt}}c@{\hspace{3pt}}|@{\hspace{3pt}}c@{\hspace{3pt}}|@{\hspace{3pt}}c@{\hspace{3pt}}|@{\hspace{3pt}}c@{\hspace{3pt}}}
        Param.   & Coef. & SE & z & $P>|z|$ & 95\% CI \\
        \hline
        $A'$        & $3.238$  & $0.786$ & $4.121$  & $0.000$ & $[1.698, 4.778]$ \\
        (intcp) & -$0.647$ & $0.297$ & -$2.176$ & $0.030$ & $[$-$1.230,$-$0.64]$ \\
    \end{tabular}}
    \end{minipage}
    }

    \vspace{-6pt}
    \caption{Bar charts showing the percentage of correctly answered trials per visualization type for the (top)~\textit{positional} accuracy and (bottom)~\textit{amplitude} accuracy. Each plot shows the $A'$ value horizontally and the percentage correct vertically. The logistic regression (in red) with a 95\% confidence interval is shown for each, with a summary table of model parameters below. Statistically significant models (i.e., $p<.05$) are in bold.}
    \label{fig:logit}
    \vspace{-7pt}
    
\end{figure*}

\renewcommand{\arraystretch}{1}

\subsection{Overall Task Accuracy} \label{sec:task_accuracy}

We begin by evaluating the accuracy of the different visualizations.

\newcolumntype{?}{!{\vrule width 1.5pt}}

\para{Positional Variation}
The results for positional variation can be seen in \autoref{table:pos_gauss_correctness_uncertainty}. To test whether the overall accuracy was statistically significant, a binomial test was conducted on each visualization, with a null hypothesis of 50\% accuracy (i.e., guessing). All visualizations \textit{except} Reeb graphs showed significance. 
To test for the significance of the differences in the overall accuracy between methods, we performed a $\chi^2$ contingency test with the null hypothesis that all methods perform identically. The result, $\chi^2(3, N=1191)=8.09,p=.044$, shows the differences are significant. Overall, color maps performed the best of all visualizations, persistence diagrams and isocontours are almost tied in terms of accuracy, and Reeb graphs were the least accurate. However, it is important to note that our concern is with the sensitivity as $A'$ increases, not the overall accuracy.

\begin{table}[!h]
    \centering
    \caption{Summary of the results for the positional variation trials. Accuracy was measured as $N_{correct}/(N_{correct}+N_{incorrect})$. Bold indicates statistical significance ($p<.05$) calculated using a binomial test.}
    \label{table:pos_gauss_correctness_uncertainty}

    \resizebox{0.985\linewidth}{!}{%
    \begin{tabular}{l|@{ }c@{ }|@{ }c@{ }|@{ }c@{ }|@{ }c ? c@{ }|@{ }c@{ }|@{ }c@{ }|@{ }c@{ }|@{ }c@{ }}
    \multicolumn{1}{@{}c@{}}{Method} & \multicolumn{4}{@{}c@{}}{Trials} & \multicolumn{5}{@{}c@{}}{Accuracy} \\
     & $N_{trials}$ & $N_{corr.}$ & $N_{incorr.}$ & $N_{unsure}$ & $A'\!=\!.15$ & $A'\!=\!.30$ & $A'\!=\!.45$ & $A'\!=\!.60$ & \textbf{Overall} \\
    \hline
    \multirow{2}{*}{Color maps} & \multirow{2}{*}{300} & \multirow{2}{*}{187} & \multirow{2}{*}{105} &  \multirow{2}{*}{8} & \multirow{2}{*}{62.3\%} & \multirow{2}{*}{61.4\%} & \multirow{2}{*}{56.6\%} &  \multirow{2}{*}{67.4\%} & \textbf{64.0\%}	\\
                            &     &     &     &    &    &     &     &    & \footnotesize($p\!<\!.001$)	\\
    \cline{2-10}
    \multirow{2}{*}{Isocontours} & \multirow{2}{*}{287} & \multirow{2}{*}{163} & \multirow{2}{*}{114} & \multirow{2}{*}{10} & \multirow{2}{*}{54.7\%} & \multirow{2}{*}{59.3\%} & \multirow{2}{*}{56.7\%} & \multirow{2}{*}{56.0\%} & \textbf{58.8\%}	\\
                            &     &     &     &    &     &     &     &    & \footnotesize($p\!=\!.002$)	\\
    \cline{2-10}
    \multirow{2}{*}{Reeb graph} & \multirow{2}{*}{307} & \multirow{2}{*}{156} & \multirow{2}{*}{141} & \multirow{2}{*}{10} & \multirow{2}{*}{42.7\%} & \multirow{2}{*}{47.4\%} & \multirow{2}{*}{55.1\%} & \multirow{2}{*}{58.9\%} & 52.5\%	\\
                            &     &     &     &    &     &     &     &    & \footnotesize($p\!=\!.208$)	\\
    \cline{2-10}
    Persistence         	& \multirow{2}{*}{331} & \multirow{2}{*}{188} & \multirow{2}{*}{137} &  \multirow{2}{*}{6} & \multirow{2}{*}{56.5\%} & \multirow{2}{*}{66.2\%} & \multirow{2}{*}{55.2\%} &  \multirow{2}{*}{50.6\%} & \textbf{57.8\%}	\\
    \hspace{8pt}diagrams    &     &     &     &    &     &     &     &    & \footnotesize($p\!=\!.002$)	\\
    \end{tabular}}
    
\end{table}

\para{Amplitude Variation}
The results for amplitude variation can be seen in \autoref{table:amp_gauss_correctness_uncertainty}. The same binomial test was conducted to verify that accuracy was statistically significant. Again, all methods \textit{except} Reeb graphs showed significance.
A $\chi^2$ contingency test was also used to compare the accuracy of the methods. The result, $\chi^2(3,N=1187)=7.58,p=.056$, shows the differences are just outside the range of significance. Nevertheless, persistence diagrams offered the best accuracy, followed by color maps, isocontours, and finally, Reeb graphs. However, we again note that our primary focus is sensitivity to changes in $A'$, not overall accuracy.

\begin{table}[!ht]
    \centering
    \caption{Summary of the results for the amplitude variation trials. Accuracy was measured as $N_{correct}/(N_{correct}+N_{incorrect})$. Bold indicates statistical significance ($p<.05$) calculated using a binomial test.}
    \label{table:amp_gauss_correctness_uncertainty}

    \resizebox{0.985\linewidth}{!}{%
    \begin{tabular}{l|@{ }c@{ }|@{ }c@{ }|@{ }c@{ }|@{ }c ? c@{ }|@{ }c@{ }|@{ }c@{ }|@{ }c@{ }|@{ }c@{ }}
    \multicolumn{1}{@{}c@{}}{Method} & \multicolumn{4}{@{}c@{}}{Trials} & \multicolumn{5}{@{}c@{}}{Accuracy} \\
     & $N_{trials}$ & $N_{corr.}$ & $N_{incorr.}$ & $N_{unsure}$ & $A'\!=\!.15$ & $A'\!=\!.30$ & $A'\!=\!.45$ & $A'\!=\!.60$ & \textbf{Overall} \\
    \hline
    \multirow{2}{*}{Color maps} & \multirow{2}{*}{311} & \multirow{2}{*}{178} & \multirow{2}{*}{122} &  \multirow{2}{*}{11} & \multirow{2}{*}{48.8\%} & \multirow{2}{*}{52.9\%} & \multirow{2}{*}{60.3\%} &  \multirow{2}{*}{67.1\%} & \textbf{59.3\%}	\\
                            &     &     &     &    &      &     &     &    & \footnotesize($p\!<\!.001$)	\\
    \cline{2-10}
    \multirow{2}{*}{Isocontours} & \multirow{2}{*}{329} & \multirow{2}{*}{177} & \multirow{2}{*}{140} & \multirow{2}{*}{12} & \multirow{2}{*}{48.7\%} & \multirow{2}{*}{49.4\%} & \multirow{2}{*}{54.9\%} & \multirow{2}{*}{62.5\%}  & \textbf{55.8\%}	\\
                            &     &     &     &    &      &     &     &    & \footnotesize($p\!=\!.022$)	\\
    \cline{2-10}
    \multirow{2}{*}{Reeb graph} & \multirow{2}{*}{304} & \multirow{2}{*}{155} & \multirow{2}{*}{142} & \multirow{2}{*}{7} & \multirow{2}{*}{52.8\%} & \multirow{2}{*}{48.1\%} & \multirow{2}{*}{48.5\%} & \multirow{2}{*}{53.9\%} & 52.2\%	\\
                            &     &     &     &    &      &     &     &    & \footnotesize($p\!=\!.243$)	\\
    \cline{2-10}
    Persistence         	& \multirow{2}{*}{279} & \multirow{2}{*}{172} & \multirow{2}{*}{101} &  \multirow{2}{*}{6} & \multirow{2}{*}{47.4\%} & \multirow{2}{*}{57.5\%} & \multirow{2}{*}{65.7\%} &  \multirow{2}{*}{81.3\%} & \textbf{63.0\%}	\\
    \hspace{8pt}diagrams    &     &     &     &    &      &     &     &    & \footnotesize($p\!<\!.001$)	\\
    \end{tabular}}  
    
\end{table}

\para{Statistics of Unsure Selections}
 Tables~\ref{table:pos_gauss_correctness_uncertainty} and \ref{table:amp_gauss_correctness_uncertainty} show a small number of unsure selections, and participants answered the questions more than 95\% of the time for all visualizations. Therefore, the results did not show any significant differences between visualization types.

\subsection{Sensitivity to Variation} \label{sec:task_sensitivity}

Here, we evaluate the sensitivity of different visualizations to changes in features (i.e., $A'$). Please refer to~\autoref{sec:measureSensitivity} describing visualization sensitivity with respect to parameter $A'$. Ideally, as $A'$ increases, the accuracy of selections should also increase for a visualization. 

\para{Positional Variation}
\autoref{fig:logit}(top) shows the bar charts and the logistic regression of positional variation accuracy for different $A'$ values for each visualization type (see \autoref{table:pos_gauss_correctness_uncertainty} for exact accuracies). For color maps and isocontours, the nearly flat line and large $p$ values indicate that they are not sensitive to increases in $A'$. Persistence diagrams show a downward trend that, although not statistically significant, suggests that higher $A'$ values were detrimental to the accuracy. Finally, the Reeb graph is the only method that shows a statistically significant upward trajectory. In other words, for positional variations, Reeb graphs are the only method sensitive to changes in $A'$.

\para{Amplitude Variation}
\autoref{fig:logit}(bottom) shows the bar charts and logistic regression for different $A'$ values of each visualization for changes in amplitude (see \autoref{table:amp_gauss_correctness_uncertainty} for exact accuracies). Color maps show a statistically significant upward trend. Isocontours also have an upward trend; however, it is slightly outside of statistical significance. Reeb graphs show a flat line and high $p$-value, which indicates that they are not sensitive to changes in $A'$. Lastly, persistence diagrams react the most to increases in $A'$, as they show the steepest statistically significant upward trend.

\subsection{Time and Interaction} \label{sec:task_time_accuracy}

We next evaluate whether using any of the visualizations resulted in longer time or more interactions from participants.

\begin{figure}[!b]
    \centering
    \subfloat[Answer time\label{fig:time_taken_box_plot}]{{\includegraphics[trim=5pt 10pt 8pt 70pt, clip, height=2.5cm]{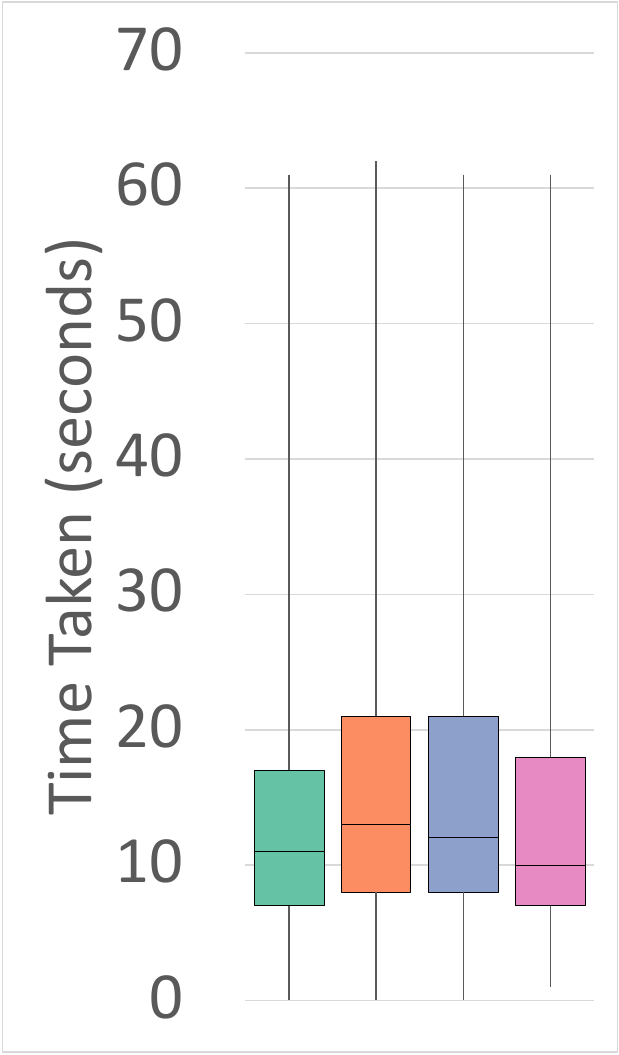}}}
    \hspace{5pt}
    \subfloat[Accuracy with interaction\label{fig:interaction_accuracy} ]{{\includegraphics[trim=5pt 10pt 5pt 70pt, clip, height=2.5cm]{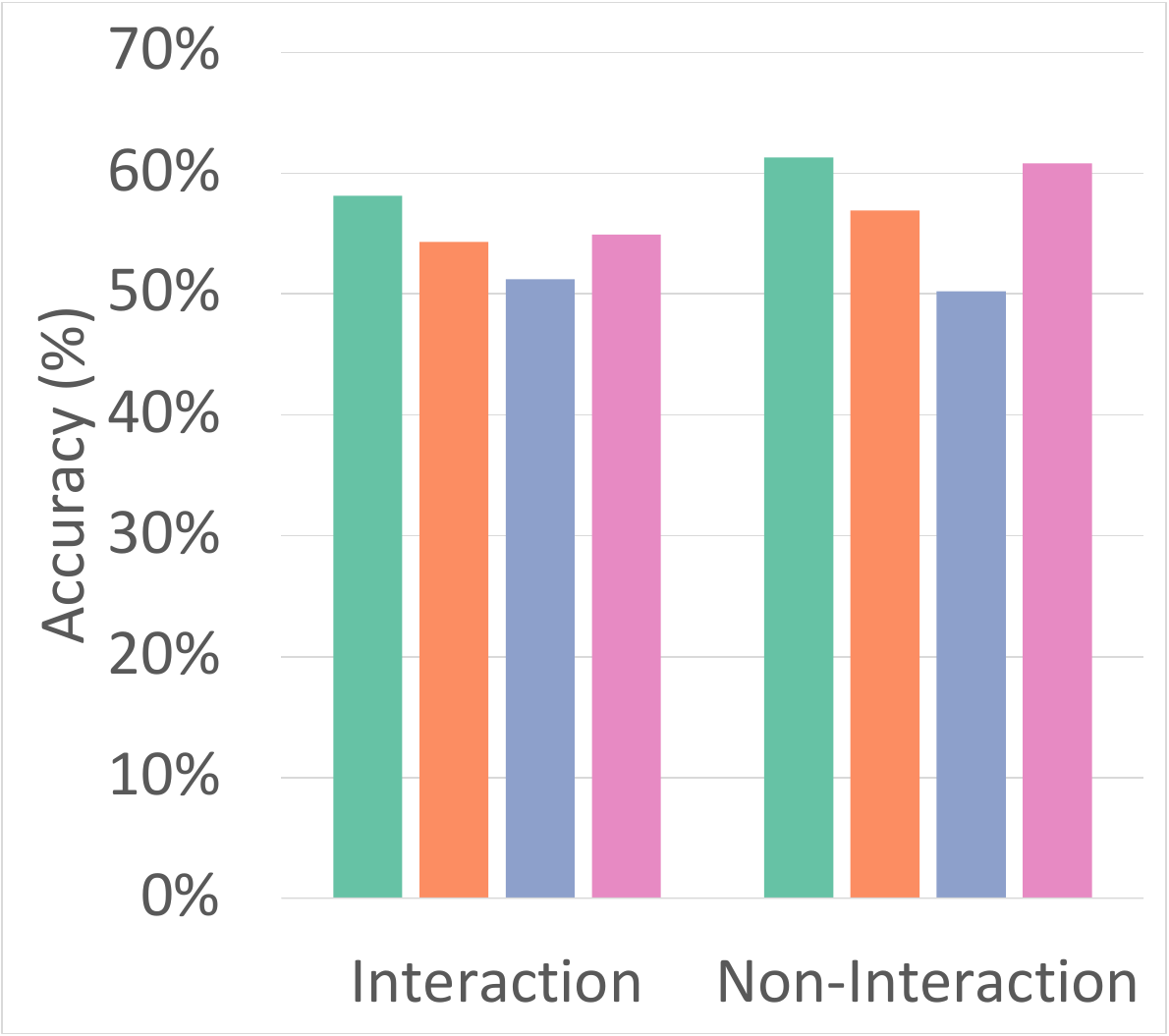}}}
    \hspace{5pt}
    \subfloat[Perceived difficulty\hspace{20pt}\label{fig:perceived_difficulty_visualization_bar_chart}]{{\includegraphics[trim=12pt 10pt 17pt 70pt, clip, height=2.5cm]{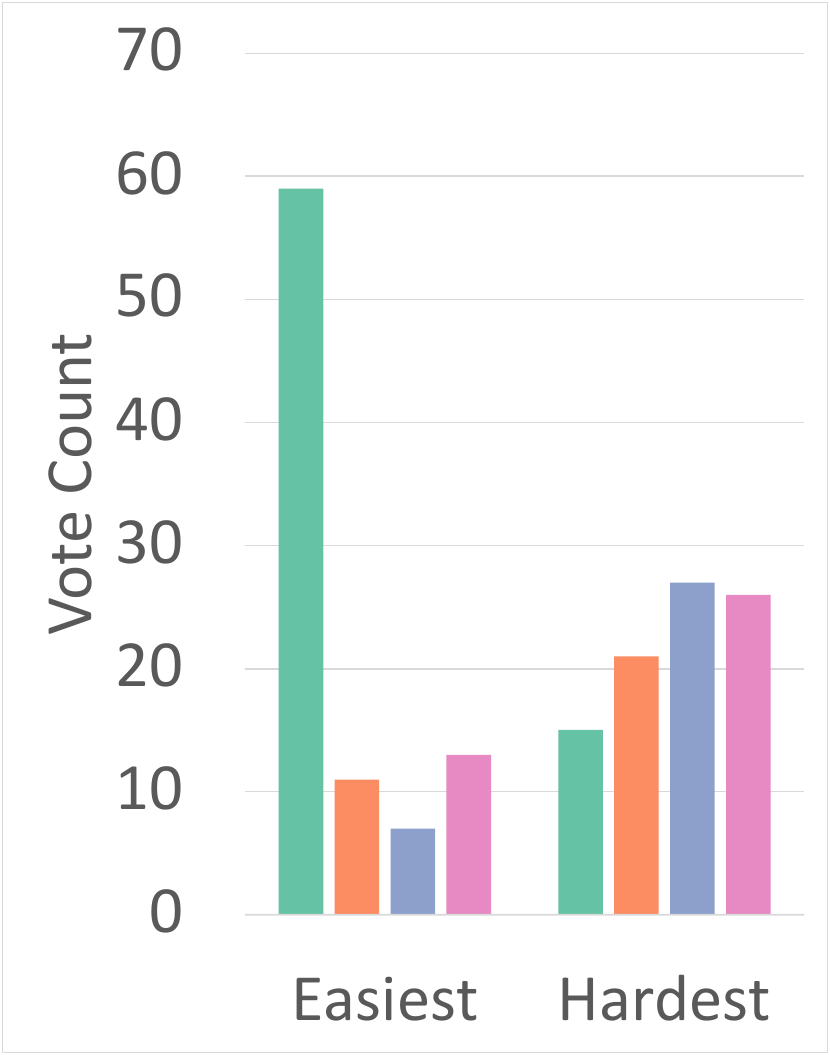}}\hspace{5pt}{\includegraphics[trim=175pt 5pt 27pt 40pt, clip, height=1.75cm]{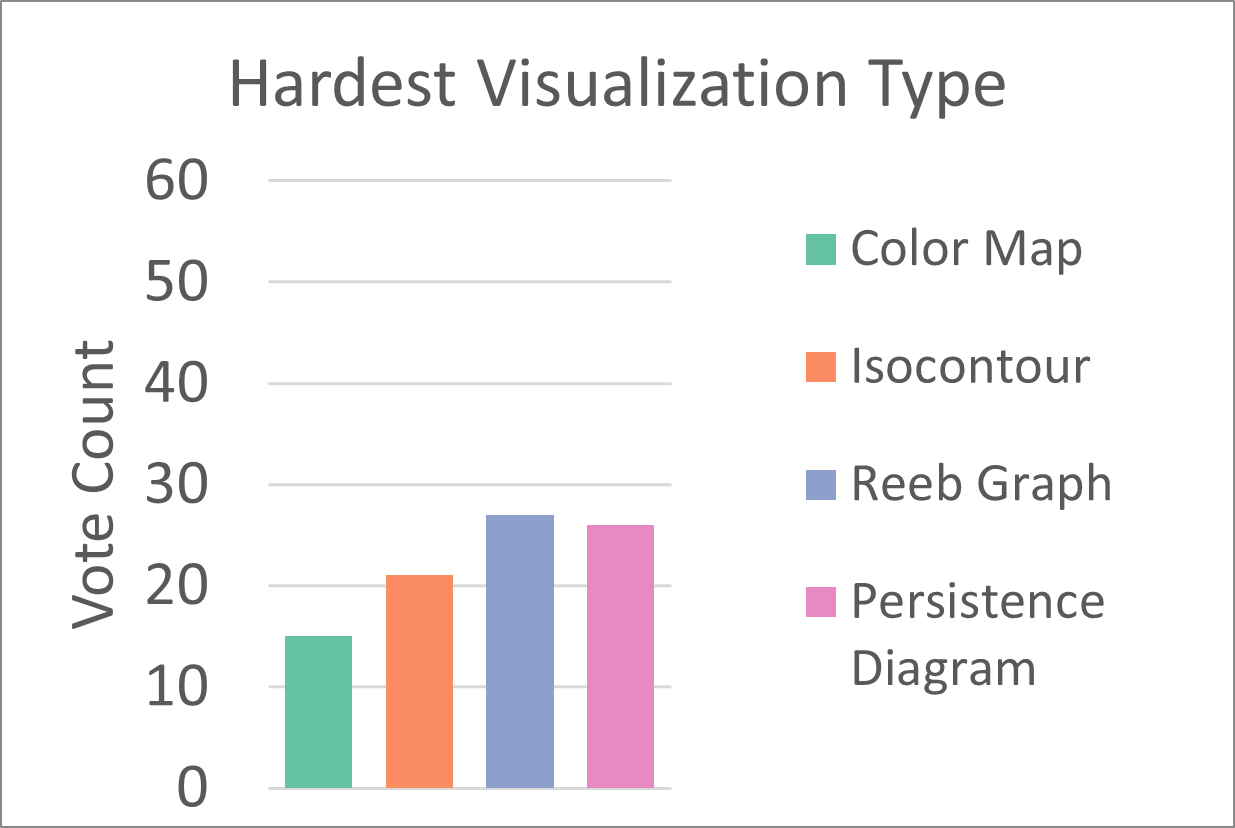}}}

    \vspace{-5pt}
    \caption{(a)~A box plot summarizing the time it took participants to answer for each visualization type. (b)~A bar chart showing the accuracy of answers with and without interactions for each visualization type. (c) Results from the post-experiment survey that asked participants what visualization type was the easiest and hardest to use.}
\end{figure}

\para{Time Taken}
The time it took for participants to answer each question was similar for each visualization, as shown in \autoref{fig:time_taken_box_plot}. We also found no clear trend between the time it took a participant to answer a question and the accuracy of their answer.

\para{Interactions}
We also evaluated accuracy when participants used some form of interaction (i.e., mouse click, mouse scroll, slider movement). We found no significant relationship between the use of interactions and the accuracy of the answers, as shown in \autoref{fig:interaction_accuracy}.

\subsection{Qualitative Feedback}\label{sec:qualitativeFeedback}

At the end of the experiment, participants had the opportunity to provide several forms of qualitative feedback. The responses provided by individual participants are documented in pp.\ 39-47 of the supplement.

\para{Perceived Difficulty of Visualization Methods}
Participants had the option to indicate which visualizations they thought were easiest and most challenging to use for the comparison task (see \autoref{fig:perceived_difficulty_visualization_bar_chart}). 
The overwhelming majority considered color maps to be the easiest. %
The results for the hardest to use were mixed, but they did indicate that Reeb graphs and persistence diagrams tended to be harder to understand, even though they had the best performance on positional and amplitude variations, respectively.

\para{Participant Feedback}
Participants were also asked to provide additional information on what approaches they used to answer the questions. The typical response for color maps was that they tried to find the variation that shared the most significant number of cold and hot areas compared to the baseline. For isocontours, they tried to infer similarities by looking at the amount of space between contour lines. Moreover, some participants also mentioned that the level-of-detail slider helped reduce the contours' density when it was hard to identify similarities. For Reeb graphs, many participants indicated that they had difficulties understanding the graph, and for some, it was helpful to reduce the level of detail of the graph. Lastly, for persistence diagrams, participants mentioned that they tried to identify common clusters of points away from the diagonal.

%% file: sec-discussion.tex
\section{Discussion \& Conclusions} 
\label{cp:conc}

\subsection{Visualization Accuracy and Sensitivity}

\para{Color Maps} Color maps performed well in our experiment. Although they showed the highest accuracy for positional variation, they also did \textit{not} show a statistically significant sensitivity to positional variation, thereby causing us to reject this hypothesis. For amplitude variation, color maps performed second best overall and showed a strong effect in terms of sensitivity, thereby allowing us to confirm that hypothesis. \textit{Overall, we can confirm color map sensitivity to amplitude variations but cannot confirm its sensitivity to positional variation.}

\para{Isocontours} The accuracy and sensitivity results for isocontours were overall in the least agreement with our expectations. In terms of accuracy, isocontours were in the middle of the pack for both variation types. Furthermore, isocontours showed no sensitivity to positional variation, thereby causing us to reject that hypothesis. For amplitude variation, we found a weak, nonstatistically significant sensitivity. Technically, this confirms our hypothesis (that isocontours have no sensitivity to amplitude changes), but we consider this result ambiguous. \textit{Overall, isocontours showed no sensitivity to positional variation and ambiguous results on amplitude variation.}

\para{Reeb Graph} Reeb graphs showed overall lower task accuracy compared to the other visualization types, which was expected behavior considering the discrete, high-frequency, and sparse nature of Reeb graphs. Although the overall accuracy for Reeb graphs was low, the Reeb graph was the only visualization type to show sensitivity to positional variations, thereby confirming this hypothesis. Interestingly, Reeb graphs performed poorly at lower $A'$ values (i.e., below 50\%, which is worse than guessing), possibly due to contradictory information generated by changes in the noise, which played an outsize role in the visualization. In addition, Reeb graphs showed no statistically significant effect in the sensitivity to variations in amplitude, thereby confirming this hypothesis as well. \textit{Overall, as predicted, Reeb graphs showed sensitivity to position variations but not amplitude variations.}

\para{Persistence Diagrams} The results for persistence diagrams showed that, as hypothesized, they were not sensitive to positional variations. Interestingly, they showed a non-statistically significant negative trend as $A'$ increased. We speculate that this is due to artifacts caused by birth-death pairing switches that occur when critical points move apart. For amplitude variation, persistence diagrams showed both the highest accuracy and sensitivity to variations in $A'$, thereby confirming this hypothesis. \textit{Overall, as predicted, persistence diagrams were sensitive to amplitude variations but were not sensitive to positional variations.}

\subsection{Implications}

There are several important implications for our findings. 

\para{Rejected Hypotheses} The several hypotheses that were rejected might be as important as those that were confirmed. Our hypotheses come from multiple decades worth of combined experience in scientific visualization. These rejected hypotheses potentially signify important misconceptions about the effectiveness of certain visualization types that further studies may illuminate.

\para{Color Maps and Isocontours} One surprising result to us was the relative strength of color maps and the weakness of isocontours. Through our analysis of accuracy and sensitivity, color maps excelled in several aspects, whereas isocontours stood out in none.

\para{Reeb Graphs and Persistence Diagrams} Reeb graphs and persistence diagrams have demonstrated very precise utility within the context of our study of positional variation for Reeb graphs and amplitude variation for persistence diagrams. This quality can be seen as both an asset and a liability because it means each will highlight variations of their supported type, whereas other variations may be lost. 

\para{No Visualization to Rule Them All} One of the most important implications of our study is that no single visualization stood out clearly with both position and amplitude variations. On one hand, this justifies using multiview visualization of the data to identify individual position and amplitude variations in data, e.g., combining color map and Reeb graph for high sensitivity to position and amplitude. On the other hand, \textit{it means that no visualization will easily identify features with both position and amplitude variations present}.

\subsection{Ecological Validity and Future Work}

To contextualize our study, we consider multiple perspectives on the ecological validity of the work.

\para{Task Relevance} The chosen task of comparing multiple variations of scalar fields is a frequent data analysis task (see \autoref{fig:topoVisComparisonRealData}). However, it does not cover the complete suite of analysis tasks one would perform with a scalar field. Further investigation is needed to more holistically evaluate the effectiveness of these visualization types.

\para{Participant Pool} Notably, our participant pool comes from the general population instead of experts in scalar field visualization. Unfortunately, the number of participants needed, made identifying enough experts difficult. The lack of expertise in the participant pool may have played a role in some of the results (e.g., lower accuracy for some methods). 
We note that expert participant pools are not without their weaknesses either. For example, expert decisions may be influenced by the familiarity bias, which puts them at risk of falling into the Dunning-Kruger effect~\cite{TA:Kurger:1999:dunningKruger}. Nevertheless, we separately evaluated the seven participants who claimed to be regular or extensive visualization users, despite the pool not being large enough to generate any statistical significance.
The results (see supplement) were not noteworthy. We also informally evaluated the research team's performance during testing and similarly found the results did not contradict the overall findings.

\para{Perception vs.\ Data Analysis} Due to the participant pool, our experiment was explicitly designed to focus on a low-level, mostly perceptual task, to make them accessible to the participant pool.
Higher-level tasks require understanding the context of the data, the meaningfulness of topological descriptors, and understanding the relationship of the visualization to the topological descriptors. These are important (but difficult) factors to measure and teach to a nonexpert participant pool.

\para{Types of Features Generated and Evaluated}
For the sake of practicality, we used Gaussian features and Perlin noise. We did not consider other feature types (e.g., anisotropic or non-Gaussian functions) or noise (e.g., salt-and-pepper noise).
In addition, the functions were limited to two simple variation types: position and amplitude. We also did not consider the mixing of multiple types of variation (i.e., both position and amplitude variation together) or other types of variations (e.g., anisotropic shape, periodic functions, etc.). Any of these variations would likely alter the results and deserve further study. Lastly, our experimental results and analysis consider topology-based visualizations of
data sampled on a 2-manifold. In the future, we plan to extend our
evaluation to variations of multiple features,
high-dimensional scalar fields, additional visualization types (e.g., planar
Reeb graphs or Morse complexes), and vector as well as tensor fields.

\para{Role of Design and Data Size and Complexity} We are cautious in extrapolating the results of our study for a given technique (e.g., color maps having uniformly higher accuracy). Our results compare specific design variations of each visualization type, and alternatives may influence their performance. For instance, a planar Reeb graph may reduce clutter but lose spatial context. Furthermore, each visualization type may respond differently in terms of data type (e.g., volumetric), size, and complexity (e.g., number of features and noise). Evaluation of these datasets presents multiple non-trivial challenges in terms of identifying a participant pool, the increasing complexity of the visualization, and the need for context/domain knowledge to perform tasks.

\para{Next Research Steps}
The approach presented in this paper lays the foundation to study the perceptual sensitivity of topological and other scientific visualizations for more complex data sets, different visualization types, design variations within the visualizations, and other analytical tasks. Particularly, the sensitivity trends presented in our evaluation serve as an important guide to the types of features that each visualization technique is better or worse at communicating. We hope the community will use the framework of study here to further investigate these questions.